\RequirePackage{fix-cm}
\documentclass{bmcart}

\usepackage[utf8]{inputenc} 

\usepackage[english]{babel}
\usepackage[utf8]{inputenc}
\usepackage{amsmath}
\usepackage{graphicx}
\usepackage[colorinlistoftodos]{todonotes}
\usepackage{chapterbib}
\usepackage{float}
\usepackage{url}
\usepackage{hyperref}
\usepackage[normalem]{ulem}
\usepackage{color}
\usepackage{comment}
\usepackage{verbatim}
\usepackage{soul}
\usepackage{subfigure}
\usepackage{url}
\usepackage{multirow}
\usepackage{booktabs}
\usepackage{natbib}

\renewcommand\tabcolsep{3pt}

\usepackage{tikz}
\usetikzlibrary{shapes,snakes}
\usepackage{amsmath,amsfonts, slashed, amssymb, wrapfig,mathtools}
\usepackage{hyperref}
\usepackage{soul}

\startlocaldefs

\endlocaldefs
\definecolor{orcidlogocol}{HTML}{A6CE39}
\definecolor{green2}{RGB}{63,142,87}
\definecolor{brown2}{RGB}{136,33,17}

\let\oldst\st
\renewcommand{\st}[1]{\textcolor{magenta}{\oldst{#1}}}

\hypersetup{colorlinks=true, citecolor=orange, urlcolor=blue, linkcolor=cyan}
\newcommand\fasa[1]{{\color{black}#1}}
\newcommand\mape[1]{{\color{black}#1}}

\begin{document}

\begin{frontmatter}

\begin{fmbox}
\dochead{Research}


\title{Flow of online misinformation during the peak of the COVID-19 pandemic in Italy}

\author[
   addressref={aff1,aff2,aff3},                   
   corref={aff1},                       
   noteref={n1},                        
   email={Guido.Caldarelli@unive.it}   
]{\inits{G}\fnm{Guido} \snm{Caldarelli}}
\author[
   addressref={aff3,aff5},
   noteref={n1},
   email={Rocco.DeNicola@imtlucca.it}
]{\inits{R}\fnm{Rocco} \snm{De Nicola}}
\author[
   addressref={aff4,aff3},
   noteref={n1},
   email={M.Petrocchi@iit.cnr.it} 
]{\inits{M}\fnm{Marinella} \snm{Petrocchi}}
\author[
   addressref={aff3},
   noteref={n1},
   email={Manuel.Pratelli@imtlucca.it}
]{\inits{M}\fnm{Manuel} \snm{Pratelli}}
\author[
   addressref={aff3},
   noteref={n1},
   email={Fabio.Saracco@imtlucca.it}
]{\inits{F}\fnm{Fabio} \snm{Saracco}}


\address[id=aff1]{%
  \orgname{Department of Molecular Sciences and Nanosystems, Ed. Alfa},
  \street{Via Torino 155},
  \postcode{30170}
  \city{Venezia Mestre},
  \cny{Italy}
}
\address[id=aff2]{%
  \orgname{European Centre for Living Technology (ECLT)},
  \street{Ca' Bottacin, 3911 Dorsoduro Calle Crosera},
  \postcode{30123}
  \city{Venice},
  \cny{Italy}
}
\address[id=aff3]{%
  \orgname{IMT Alti Studi Lucca},           
  \street{Piazza San Francesco 19},                 
  \postcode{55100}                              
  \city{Lucca},                             
  \cny{Italy}                                   
}
\address[id=aff4]{%
  \orgname{Institute of Informatics and Telematics, National Research Council},
  \street{via Moruzzi 1},
  \postcode{56124}
  \city{Pisa},
  \cny{Italy}
}
\address[id=aff5]{%
  \orgname{CINI - National Laboratory for Cybersecurity},
  \street{via Ariosto, 25},
  \postcode{00185}
  \city{Roma},
  \cny{Italy}
}

\begin{artnotes}
\note[id=n1]{Equal contributor} 
\end{artnotes}

\end{fmbox}


\begin{abstractbox}

\begin{abstract} 
 \fasa{The COVID-19 pandemic has impacted on every human activity and, because of the urgency of finding the proper responses to such an unprecedented emergency, it generated a diffused societal debate. The online version of this discussion was not exempted by the presence of d/misinformation campaigns, but differently from what already witnessed in other debates, the COVID-19 -intentional or not- flow of  false information put at severe risk the public health, reducing the effectiveness of governments' countermeasures. In the present manuscript, we study the \emph{effective} impact of misinformation in the Italian societal debate on Twitter during the pandemic, focusing on the various discursive communities.

In order to extract the discursive communities, we focus on verified users, i.e. accounts whose identity is officially certified by Twitter. 

We thus infer the various discursive communities based on how verified users are perceived by standard ones: if two verified accounts are considered as similar by non unverified ones, we link them in the network of certified accounts.
We first observe that, beside being a mostly scientific subject, the COVID-19 discussion show a clear division in what results to be different political groups.
At this point, by using a commonly available fact-checking software (\fasa{NewsGuard}), we assess the reputation of the pieces of news exchanged. 
\fasa{We filter the network of retweets (i.e. users re-broadcasting the same elementary piece of information, or {\em tweet}) from random noise and check the presence of messages displaying an url. The impact of misinformation posts reaches the 22.1\% in the right and center-right wing community and its contribution is even stronger in absolute numbers, due to the activity of this group: 96\% of all non reputable urls shared by political groups come from this community.}  
}
\end{abstract}

\end{abstractbox}
\section{Introduction}
\label{sec:intro}

\fasa{The advent of the internet and online social media has promoted a more democratic access to information, increasing the offer of news sources, with a significant number of individual contributions too. Unfortunately, unmediated communication channels have generated an incredible amount of low-quality contents, polluting the online debate in several areas, like politics, healthcare, education, and environment~\cite{Bradshaw2018b}.}

\fasa{For this reason, in the recent Joint Communication titled ``Tackling COVID-19 disinformation - Getting the facts right" (June 10, 2020, \url{https://bit.ly/35C1dGs}), the High Representative of the Union for Foreign Affairs and Security Policy, while introducing the various d/misinformation campaigns that arose during the first months of the COVID-19 pandemic, presented an explicit declaration of intent: 
``Combatting the flow of disinformation, misinformation [...] calls for action through the EU's existing tools, as well as with Member States' competent authorities [...] enhancing citizens' resilience."} 

\fasa{The research regarding online social media, featuring the detection of misinformation campaigns and of the pollution of the online political debate has been the target of a great flow of recent research, e.g.~\cite{Gonzalez-Bailon2013,cresci2015fame, Stella2018,Ciampaglia2018,Ferrara2016rise,Ferrara2019,Cresci2019WebSci,Bovet2019, Becatti2019, Caldarelli2020a}. Nevertheless, due to the societal relevance of the topic, the analysis of misinformation campaigns during the COVID-19 pandemic has immediately attracted several scholars, focusing on different facets of this phenomenon: on the Google trend related to Coronavirus arguments~\cite{Rovetta2020}, on the existence of Facebook groups experiencing an extreme exposure to disinformation~\cite{Celestini2020}, on the evolution of the diffusion in Twitter of false information across several countries~\cite{Gallotti2020} and on the {\em disinformation epidemiology} on various online social platforms~\cite{Cinelli2020}. 
In} the present paper, using Twitter as a benchmark, we shall consider the \fasa{\emph{effective} flow of online misinformation} in Italy, one of the \fasa{countries in Europe that have been affected the most by the COVID-19 pandemic}\footnote{In Italy, since the beginning of the pandemic and at time of writing, more than 2.5 million persons have contracted the Covid-19 virus: of these, more than 89k have died. Source: \url{http://www.protezionecivile.gov.it/} \fasa{Accessed February 4, 2021.}}\fasa{, and how this flow affected the various discursive communities, i.e., groups of users that debate on the pandemic. Since the debate is mostly centered on verified users, i.e., users whose identity is certified by Twitter, we start considering their interactions with  unverified accounts. Following~\cite{Becatti2019, Caldarelli2020a,Radicioni2020}, our intuition is that two verified users that are perceived to be similar by unverified users, interact  with \fasa{(i.e., retweet and are retweeted by)} the same accounts. In order to assess how many common unverified users are `enough' to state that the two verified users  are indeed similar, we use an entropy-based null-model as a benchmark~\cite{Squartinia,Cimini2018}. In a nutshell, the entropy-based null-model is a network benchmark in which part of the information is constrained to the values observed in the real system and the rest is completely random. If the observations are not compatible with the null-model, then they cannot be explained by the constraints only and carry a non trivial information regarding the real system.\\ Interestingly enough, we find that the main discursive communities are political, i.e., they involve politicians, political parties and journalists supporting a specific political ideal.} While, at first sight, this may sound surprising - the pandemic debate was more on a scientific than on a political ground, at least in the very first phase of its abrupt diffusion -, it might be due to pre-existing {\it echo chambers}~\cite{echo2016}. 
We then consider the news sources shared among the accounts of the various groups. With a hybrid annotation approach, based on independent journalists and annotation carried out by members of our team, we  categorised such sources as reputable and non reputable (in terms of credibility of the published news and the transparency of the source). 

\fasa{Finally, we extract the effective flow of content shared within the network: still following the approach of Ref.~\cite{Becatti2019,Caldarelli2020a}, we extend the  entropy-based methodology to a directed bipartite network of users and posts. In this sense, we are able to control not only the authorship activity and the retweeting attitude of the various accounts, but even the \emph{virality} of the different messages, i.e., how many times a single message is shared.}

\fasa{The various political groups display different online behaviours.} In particular, the right wing community is more numerous and more active, even relatively to the number of accounts involved, than the other communities.  Interestingly enough, newly formed political parties, as the one of the former Italian prime Minister Matteo Renzi, quickly imposed their presence on Twitter and on the online political debate, with a strong activity. Furthermore,  the different political parties use different sources for getting information on the spreading on the pandemic. Notably, we experience that \fasa{right and center-right wing} accounts spread information from non reputable sources with a frequency almost \fasa{10} times higher than that of the other political groups. \fasa{Interestingly, due to their outstanding activity, their impact, in terms of number of d/misinforming posts in the debate, is much greater than that of any other groups. 

The paper is organised as follows:
we describe the dataset in~Section~\ref{sec:dataset} and present the results of our analysis in Section~\ref{sec:res}. After discussing and commenting our results in Section~\ref{sec:conc}, we introduce  the  methodology implemented in our analysis (Section~\ref{sec:bg}).} 

\fasa{\section{Related Work}}
\label{sec:relwork}
\fasa{As in any disaster, natural or otherwise, people is exposed  to the spread of related online misinformation. This is the case of the COVID-19: the physical pandemic was quickly complemented by the so-called COVID-19 infodemic, i.e. the diffusion of a great amount of low-quality information regarding the pandemic.
Academia has stepped up its efforts to combat this infodemic. Here, we briefly review some of the most relevant articles in the area.

Rovetta et al., in \cite{Rovetta2020}, explore the internet search activity related to COVID-19 from January to March 2020, to analyse article titles  from the most read newspapers and government websites `to investigate the attitudes of infodemic monikers circulating across various regions and cities in Italy'. The study reveals a growing regional and population-level interest in COVID-19 in Italy, highlighting how the majority of searches concern -often unfounded- remedies against the disease. 

Work in~\cite{Gallotti2020}, by Gallotti et al., develops an Infodemic Risk Index to depict the risk of exposure to false online information in various countries around the world. Regarding healthcare news, the authors find that even before the rise of the pandemic, entire countries were exposed to false stories that can severely threaten the public health. 

Hossaini et al.~\cite{hossain-etal-2020-covidlies} release COVIDLies,  a dataset of 6761 expert-annotated tweets to evaluate the performances of existing NLP systems in detecting false stories about  COVID-19. 
Still regarding datasets, work by Zhou et al.~\cite{Zhou_2020} present ReCOVery, a repository  of more than 2k news articles on Coronavirus, together with more than 140k tweets testifying the spreading of such articles on Twitter.  
Chen et al., in~\cite{Chen_2020}, present to the scientific community a multilingual COVID-19 Twitter dataset that they have been continuously collecting since January 2020.  
Celestini et al., in~\cite{Celestini2020}, collect and analyse over 1.5 M COVID-19-related posts in Italian language. Findings are that, although controversial topics associated to the origin of the virus circulate online, discussions on such topics is negligible compared to those on mainstream news websites.

Pierri et al, in~\cite{pierri2021vaccinitaly}, provide public access to  online conversations of Italian users around vaccines on Twitter, an on-going collection capturing the Italian vaccine roll-out (on December 27, 2020).  The authors also report a consistent amount of low-credibility information already circulating on Twitter alongside vaccine-related conversations. 
Still regarding the COVID-19 vaccination campaigns, De Verna et al. collect a Twitter dataset of English posts, giving statistics about hashtags, urls, and number of tweets over time through a dashboard. 

Sharma et al, in~\cite{sharma2020identifying}, consider the role of Twitter bots in the pandemic online debate. By moving away from the research trend  of detecting teams of bots on the basis of features concerning coordination and synchronous behavior between such accounts, they propose an approach to automatically uncover coordinated group behaviours from account activities and interactions between accounts, based on temporal point processes.

A lot of works examine Twitter, because of its provision of public APIs for data taking. Yang et al.~\cite{yang2020covid19}  analyse and compare instead the presence of links pointing to low-credibility content both on Twitter and Facebook posts. Misinformation `superspreaders' and evidences of coordinated sharing of false stories about COVID-19 are present on both the platforms. Still at a narrower granularity, Cinelli et al. in ~\cite{Cinelli2020} carry on a massive analysis on  Twitter, Instagram, YouTube, Reddit and Gab. The authors characterize COVID-19-related information spreading from questionable sources, finding different volumes of misinformation in each platform.

This brief literature overview on the COVID-19 infodemic, although not exhaustive, highlights  that the spread of misinformation on pandemic-related issues on the internet and social media is a major issue. Scientists propose various  methods to detect false information about the virus. Aligned with this line of research, in this manuscript we quantify the \emph{effective} level of misinformation about the pandemic exchanged on Twitter during late winter and early spring in 2020 in Italy, with a special focus on the role of the Italian political communities.}

\section{Dataset}
\begin{table}[ht!]
\centering
\begin{tabular}{l}
Keywords and Hashtags\\
\hline
\hline
coronavirus\\
ncov\\
covid\\
SARS-CoV2\\
\#coronavirus\\
\#coronaviruses\\
\#WuhanCoronavirus\\
\#CoronavirusOutbreak\\
\#coronaviruschina\\
\#coronaviruswuhan\\
\#ChinaCoronaVirus\\
\#nCoV\\
\#ChinaWuHan\\
\#nCoV2020\\
\#nCov2019\\
\#covid2019\\
\#covid-19\\
\#SARS\_CoV\_2\\
\#SARSCoV2\\
\#COVID19\\
\hline
\end{tabular}
\smallskip
\caption{Keywords and Hashtags which drove the data collection phase\label{table:keywords}}

\end{table}
\label{sec:dataset}
\fasa{Using Twitter's streaming API from February $21^{st}$ to April $20^{th}$ 2020\footnote{We had an interruption of one day and $4$ hours on February $27^{th}$ and another of three days and $8$ hours on March $10^{th}$ due to a connection breakdown.
\fasa{Because of the validation procedure that, moreover, we applied on the aggregated network over the entire period of data collection, we expect that the effect of the breakdown to be limited for the interpretation of our results.}} we collected circa $4.5$M tweets in  Italian language.} 
\fasa{Actually, the data set analysed is a subset of a greater corpus in which the language was not a selection criterion for the download; we then selected Italian messages. Due to the great amount of data downloaded, we run into Twitter download limit, even if quite seldom\footnote{\fasa{During the peaks of traffic during the end of February we run into Twitter limits less than once a day.}}; nevertheless, due to the validation procedures (see Sections~\ref{sec:res} and~\ref{sec:bg}), we expect that the impact of Twitter limitation to be negligible in terms of the interpretation of the results.}\\
The data collection was keyword-based, with keywords related the Covid-19 pandemics, \fasa{including the most used version (\emph{coronavirus}) and the first names for the virus. The complete lists of keywords used can be found in Table~\ref{table:keywords}.}  
Twitter's streaming API returns any tweet containing the keyword(s) in the text of the tweet, as well as in its metadata. It is worth noting that it is not always necessary to have each permutation of a specific keyword in the tracking list. For example, the keyword `Covid' will return tweets that contain both `Covid19' and `Covid-19'. Table~\ref{table:keywords} lists a subset of the considered keywords and hashtags.
There are some hashtags that overlap due to the fact that an included keyword is a sub-string of another one, but we included both for completeness.\\

\fasa{Let us conclude this short section with few final comments. First, we remark that, beside being relatively less popular than other OSNs (in Italy it is used by less than 5.8\% of the population in order to access to the latest information~\cite{AGCOM2017}), Twitter has an incidence of journalists and politicians higher than other platforms (Twitter is the second most used social network, after Facebook, with an incidence of 30\% of journalists accessing it every day~\cite{AGCOM2018}), probably to the limited number of characters of the messages shared, which is extremely suitable for short and fast communication, as the breaking news~\cite{AGCOM2018}.} 
\fasa{Finally, d}etails about the \fasa{health} situation in Italy during the period of data collection can be found in the Supplementary Material, Section 1.1: `Evolution of the Covid-19 pandemics in Italy'. 


\section{Results}
\label{sec:res}
\subsection{Discursive communities of verified users}
\label{sec:polar}
\fasa{Many results in the analysis of online social networks (OSN) shows that users are highly clustered in group of opinions~\cite{Adamic2005, Conover2011, Conover2011a, Conover2012,DelVicario2016c, DelVicario2017a,Quattrociocchi2014,Zollo2017,Zollo2015}; indeed those groups have some peculiar behaviours, as the echo chamber~\cite{DelVicario2016c, DelVicario2017a}. Following the example of references~\cite{Becatti2019, Caldarelli2020a}, we leverage this users' clustering in order to detect discursive community, i.e. groups of accounts interacting among themselves by retweeting on the same (covid-related) subjects. Remarkably, our procedure does not follow the analysis of the text shared by the various users, but is simply related on the retweeting activity among users. In the present subsection we will examine how the information about discursive community of verified Twitter users can be extracted. \\ 
On Twitter there are two distinct categories of accounts: verified and unverified users. Verified users have a tick close to the screen name; the platform itself, upon request from the user, has a procedure to check the authenticity of the account. Verified accounts are usually owned by politicians, journalists or VIPs in general, as well as ministers, newspapers, newscasts, companies and so on: for those kind of users, the verification procedure guarantees the identity of their account and reduce the risk of malicious accounts tweeting in their name. Non verified accounts are for \emph{standard} users: in this second case, we cannot trust any information provided by the users. The information carried by verified accounts has been studied extensively in order to have a sort of anchor for the related discussion~\cite{Hentschel2014,varol2019verified,Bovet2019, Becatti2019, Caldarelli2020a,Radicioni2020}\\
To detect the discursive communities we consider the bipartite network represented by verified (on one layer) and unverified (on the other layer) accounts: a link is connecting the verified user $v$ with the unverified one $u$ if at least one time $v$ was retweeted by $u$, and/or viceversa. To extract the similarity of users, we compare the commonalities with a bipartite entropy-based null-model, the Bipartite Configuration Model (\emph{BiCM}~\cite{Saracco2015a}), described in details in subsection~\ref{ssec:bicm}. 
The {\it rationale} is that two verified users that share  many links to same unverified accounts probably have similar visions, as perceived by the audience of unverified accounts. We then apply the method of~\cite{Saracco2016}, in order to get a statistically validated projection of the bipartite network of verified and unverified users. In a nutshell, the idea is to compare the amount of common linkage measured on the real network with the expectations of an entropy-based null model fixing (on average) the degree sequence: if the associated p-value is so low that the overlaps cannot be explained by the model, i.e. such that it is not compatible with the degree sequence expectations, they carry non trivial information and we put a link connecting the two nodes in the (monopartite) projection of verified users.}\\

The \fasa{top} panel of Fig.~\ref{fig:subcomm_netwk}
shows the network obtained by following the above procedure. The network resulting from the projection procedure will be called, in the rest of the paper, \textit{validated} network\footnote{\fasa{The term \emph{validated} should not be confused with the term \textit{verified}, which instead denotes a Twitter user who has passed the formal authentication procedure by the social platform.}}.

In order to get the community of verified Twitter users, we applied the Louvain algorithm~\cite{Blondel2008} to the data in the \fasa{undirected} validated network.  Such an algorithm, despite being one of the most \fasa{effective and} popular, is also known to be order dependent~\cite{Fortunato2010}. To get rid of this bias, we apply it iteratively $N$ times ($N$ being the number of the nodes) after reshuffling the order of the nodes. Finally, we select the partition with the highest modularity. The network presents a strong community structure, composed by four main subgraphs. 
When analysing 

the emerging 4 communities, we find that they correspond to   \begin{enumerate}
    \item \fasa{Media and right and center-right wing parties} (in steel blue)
    \item Center-left wing (in dark red)
    \item Movimento 5 Stelle (\emph{5 Stars Movement}, or M5S; in dark orange)
    \item Institutional accounts (in sky blue) 
\end{enumerate}  
Details about the political situation in Italy during the period of data collection can be found in the Supplementary Material, Section 1.2: `Italian political situation during the Covid-19 pandemic'. 

\fasa{While the various groups display a quite evident homophily among their components, we further examined them by re-running the Louvain algorihtm inside each of them, with the same care as above to the node order.}

\begin{figure}[ht!]
\includegraphics[width=.6\linewidth]{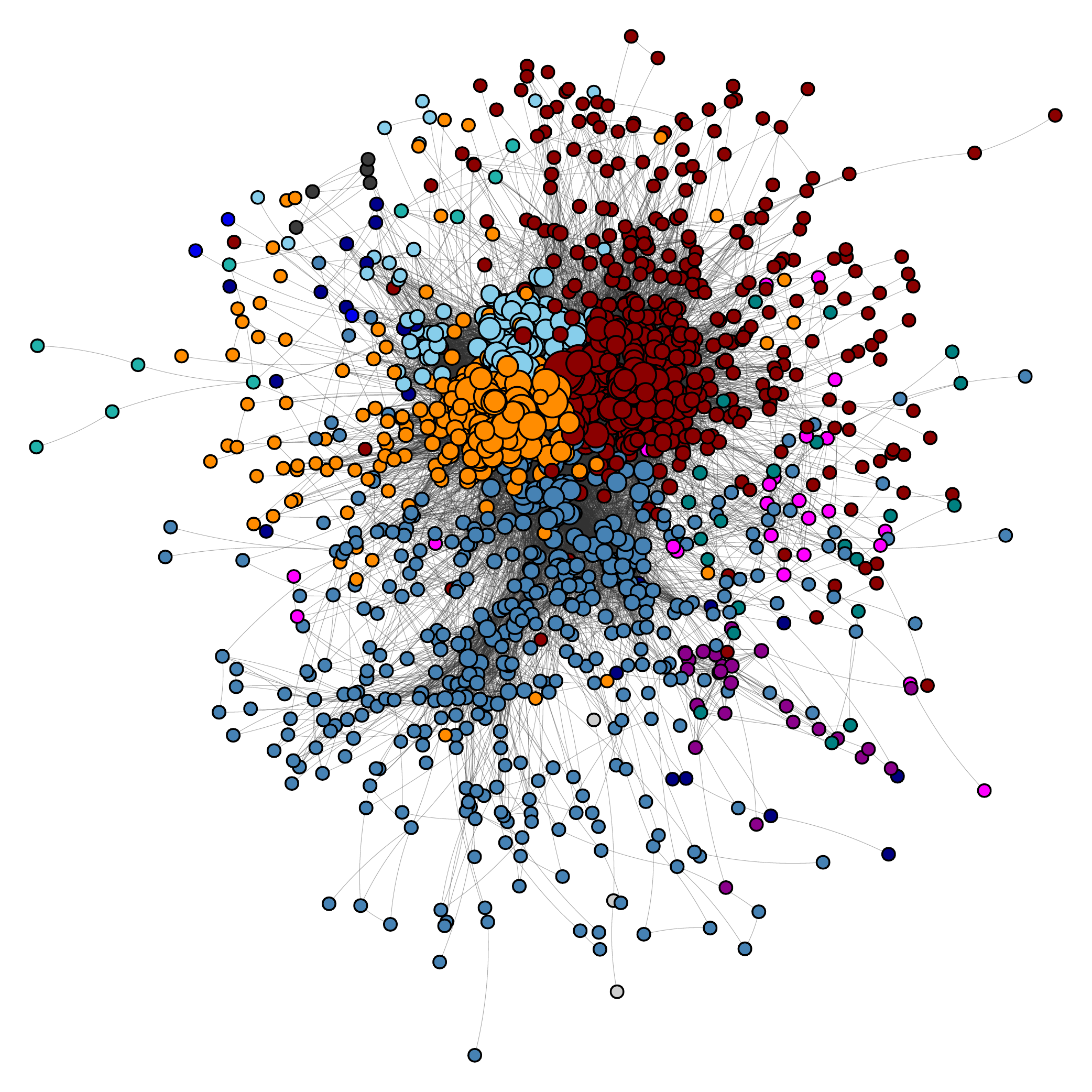} 
\includegraphics[width=.6\linewidth]{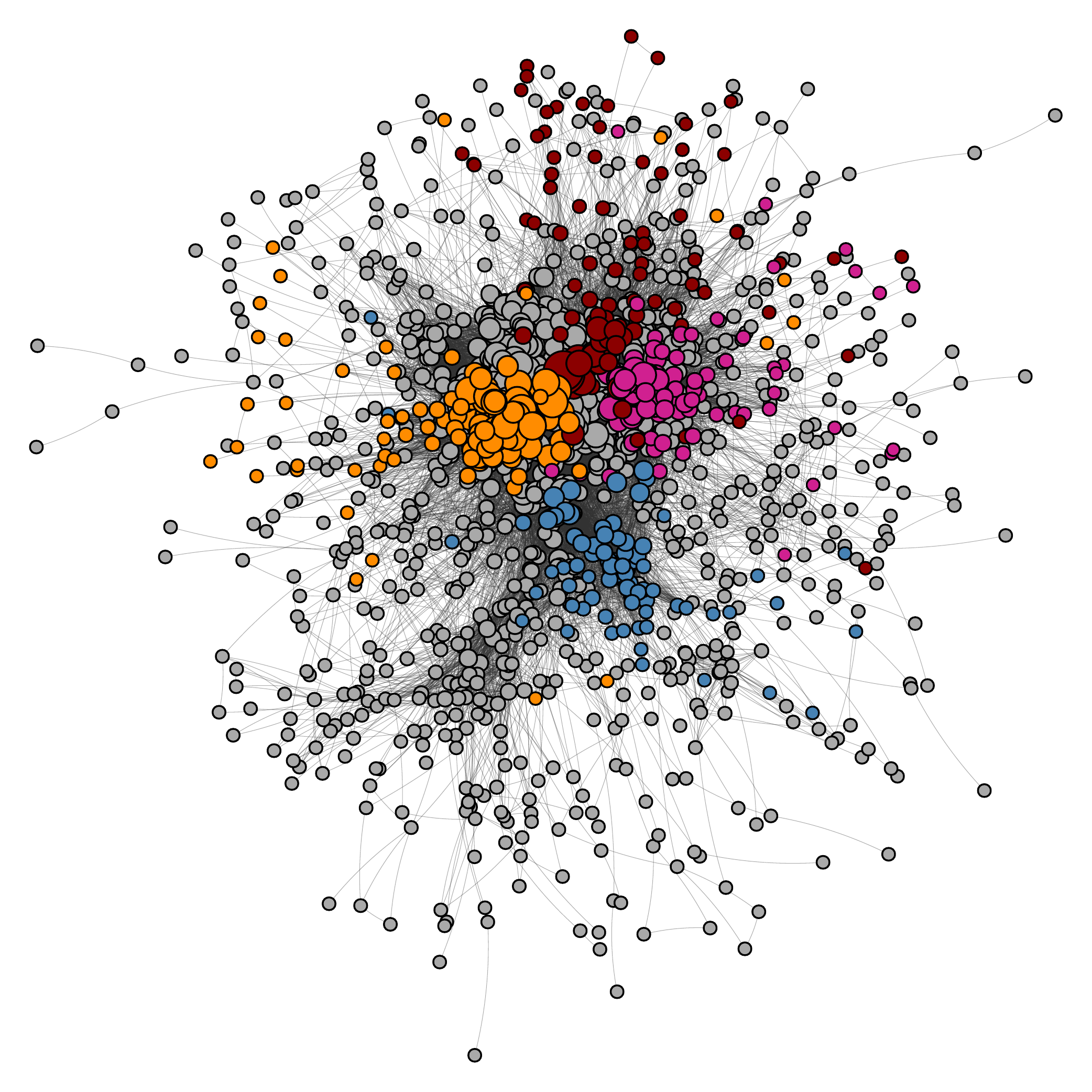}
    \caption{\fasa{\textbf{Discursive communities of verified users:} on the left communities, found running a Louvain community detection algorithm on the Largest Connected Component (LCC) of the validated network of verified users. In red, in the top right corner, center-left wing parties; in sky blue (on top) the official government accounts; in orange the Movimento 5 Stelle oriented community and in steel blue (on the bottom) the news media and center-right and right wing community. Other minor communities can be found in the periphery of the LCC. Actually, by rerunning the same community detection inside each of the largest observed communities, it is possible to find \emph{purely} political subcommunities, i.e. communities composed quite exclusively by politicians and official accounts of political parties. In the right panel, they are highlighted: in magenta, Italia Viva, the political party of the former prime Minister Matteo Renzi; in red the Partito Democratico, i.e. the Italian Democratic Party; in orange the Movimento 5 Stelle (\emph{5 Stars Movement}, or M5S) and in blue the center-right and right wing parties Forza Italia, Lega and Fratelli d'Italia. A more detailed description of the subcommunities of the network can be found in the section 2 of the Supplementary Material.} In both panels the node dimensions are proportional to their degree.}
    \label{fig:subcomm_netwk}
\end{figure}

\fasa{Since the subcommunities structure is extremely rich, we suggest the interested reader to consult the section 2 of the Supplementary Material for a more detailed description. In this manuscript we focus on the purely political subcommunities, highlighted in the lower panel of Fig.~\ref{fig:subcomm_netwk}.
Starting from the center-left wing, we can find a darker red community, including  the main politicians of the Italian Democratic Party (\emph{Partito Democratico}, or PD), as well as by its representatives in the European Parliament (Italian and others) and some EU commissioners.} The violet red group is instead mostly composed by the representatives of Italia Viva, a new party founded by the former Italian prime minister Matteo Renzi (December 2014 - February 2016). 

In turn, also the dark orange (M5S) community shows \fasa{the presence of a purely political} subcommunity (in orange in the bottom panel of Fig.~\ref{fig:user_polarization_network}), which contains the accounts of politicians, parliament representatives and ministers of the 5 Stars Movement and journalists \fasa{of \emph{Il Fatto Quotidiano}, a newspaper supporting the Movimento 5 Stelle.}

Similar considerations apply to the steel blue community: the subcommunity of center-right and right wing parties (as Forza Italia, Lega and Fratelli d'Italia\fasa{, from now on FI-L-FdI) is represented in blue in the bottom panel of Fig.~\ref{fig:user_polarization_network}}. 

Finally, the sky blue community is mainly composed by Italian embassies around the world.\\ 
\fasa{Let us conclude with a final comment.
In~\cite{Caldarelli2020a}, the authors analysed  with similar techniques the Twitter Italian debate on a political subject as the migration policies. After cleaning the system from random noise, as in the present paper, the authors highlighted a group of coordinated accounts -the \emph{bot squad}- increasing the visibility of more than a single genuine account. Beside the different final target, the division in community resembles the one found here, with few differences. First, in~\cite{Caldarelli2020a} media and center-right and right wing parties appeared in different communities from the very beginning; this is probably due to the fact that in the present case, the criticisms regarding the management of the pandemic by the main leaders of these parties were promptly reported by media, since they represented the opposition to the government. Secondly, in Ref.~\cite{Caldarelli2020a} M5S is not distinguishable from the the right and center-right wing discursive community. This second point is not so surprising, since at the time of the data collection of the previous manuscript, M5S was governing in an alliance with Lega, the main right wing party in Italy, and Matteo Salvini, the leader of Lega, was the Minister of Internal Affairs. In this sense, from the data, M5S appeared to share the opinions of its ally on the specific argument of migration policies. Due to the different subject and to the different political scenario (at the moment of the data collection, M5S, PD and Italia Viva are at the government: the interested reader can find more information about the Italian political situation in the Supplementary Material, in Subsection 1.2), in the present analysis M5S manifests its individuality.
}

\fasa{\subsection{Domain names' analysis}}
\label{subsec:domain-verified}

Here, we report a series of analyses related to the domain names, hereafter simply called domains, that mostly appear in all the tweets of the validated network of verified users. The domains have been tagged according to their degree of credibility and transparency,  as indicated by the independent software toolkit NewsGuard \url{https://www.newsguardtech.com/}. The details of this procedure are reported below.

As a first step, we considered the network of verified accounts, whose communities and subcommunities are shown
in Fig.~\ref{fig:subcomm_netwk}. On this topology, we labelled all domains that had been shared at least 20 times (between tweets and retweets). 
\begin{table}[ht!]
\centering
\begin{tabular}{c|l}
label & \text{description}\\
\hline
\hline
R & Reputable news source\\
$\sim\text{R}$ & Quasi Reputable news source\\
NR & Not Reputable news source\\
S & social network\\
F & fundraiser and petition site\\
M & marketplace\\
P & official journal of a political party\\
IS & institutional site\\
ST & online streaming platform\\
SE & search engine\\
UNC & unclassified\\
\hline
\end{tabular}

\smallskip
\caption{Tags used for labeling the domains\label{table:domains-tags}}
\end{table}

Table~\ref{table:domains-tags} shows the tags associated to the domains. In the rest of the paper, we shall be interested in quantifying reliability of news sources publishing during the period of interest. 
Thus, for our analysis, we will not consider those sources corresponding  to  social networks, marketplaces, search engines, institutional sites, etc.\fasa{; nevertheless, the information regarding their frequency are available to the interested readers in the Supplementary Material.} Tags R, $\sim\text{R}$ and NR in Table~\ref{table:domains-tags} are used only for news sites, be them newspapers, magazines, TV or radio social channels, and they stand for Reputable,  Quasi Reputable, Not Reputable,  respectively.

As mentioned above, we relied on NewsGuard, a plugin resulting from the joint effort of journalists and software developers aiming at evaluating news sites according to nine criteria concerning credibility and transparency. For evaluating the credibility level, the metrics consider, e.g., whether the news source regularly publishes false news, does not distinguish between facts and opinions, does not correct a wrongly reported news. For transparency, instead, the tool takes into account, e.g., whether owners, founders or authors of the news source are publicly known; and whether advertisements are easily recognizable.
After combining the individual scores obtained out of the nine criteria, the plugin associates to a news source a score from 1 to 100, where 60 is the minimum score for the source to be considered reliable.  When reporting the results, the plugin  provides details about the criteria which passed the test and those that did not. \mape{For the sake of completeness, the Supplementary Material reports the procedure adopted by Newsguard journalists and editors to score each news site, the meaning of the score, and which are the textual information associated with the score. The material is inherited from the Newsguard website\footnote{NewsGuard rating process: \url{https://www.newsguardtech.com/ratings/rating-process-criteria/}}}.

In order to have a sort of no-man's land and not to be too abrupt in the transition between reputability and non-reputability, when the score was between 55 and 65, we considered the source to be quasi reputable, $\sim$R.

It is worth noting that not all the domains in the dataset  under investigation were evaluated by NewsGuard at the time of our analysis. \mape{For those not yet evaluated by Newsguard, the annotation was made by three members of our team, who assessed the domains by using a subset of the NewsGuard criteria. The final class has been decided by majority voting (it never happened that the three annotators gave 3 different labels to the same domain). In the case of the network of verified users, considering only domains that appear at least 20 times, we have 80 domains annotated by Newsguard and 42 domains annotated by our three annotators. We computed the Fleiss' kappa ($\kappa$) inter-rater agreement metric~\cite{gwet2014}. The metric measures the level of agreement of different annotators on a task.  The annotators showed a moderate agreement for the classification of domains, with $\kappa = 0.63$. 
}

Table~\ref{table:verified_global_info} gives statistics about number and kind of tweets, the number of url and distinct url (dist url), the number of domains and users in the validated network of verified users. We clarify what we mean by these terms with an example: a domain for us corresponds to the so-called `second-level domain' name\footnote{\url{https://en.wikipedia.org/wiki/Domain_name}}, i.e., the name directly to the left of .com, .net, and any other top-level domains. For instance, \url{repubblica.it}, \url{corriere.it}, \url{nytimes.com} are considered domains by us. Instead, the url maintains here its standard definition\footnote{\url{https://en.wikipedia.org/wiki/URL}} and an example is \url{http://www.example.com/index.html}.

Table~\ref{table:verified_global} shows the outcome of the domains annotation, according to the scores of NewsGuard or to those assigned by the three annotators, when scores were no available from NewsGuard.

 \begin{table}[ht]
\centering
\fasa{
\begin{tabular}{cccccc}
type & \#post & \#url & \#dist url & \#domain & \#user\\
\hline
\hline

tw & 46277 & 37095 & 32605 & 1168 & 1115 \\
rt & 17190 & 9796 & 7504 & 1178 & 1385 \\
\hline
\end{tabular}
\smallskip
\caption{\textbf{Posts, urls, domains and users statistics in the validated network of verified users.} ``Tw" represent pure tweets, while ``rt" indicates retweets. The number of tweets sharing an url is much higher than the one of retweets and it is a known results for verified users, from which they appear to drive the online debate.} 
\label{table:verified_global_info}}
\end{table}

\begin{table}[ht]
\centering
\fasa{
\begin{tabular}{cc|cccc}
type & \#url & \text{R} & $\sim$\text{R} & \text{NR} & \text{Others}\\
\hline
\hline
tw & 37095 & 71.0 & 1.1 & 1.7 & 26.2\\
rt & 9796 & 57.9 & 1.2 & 1.1 & 39.8\\
\hline
\end{tabular}
\smallskip
\caption{\textbf{Annotation results over all the domains in the validated network of verified users.} It is worth noticing that, while the original posts are sharing reputable domains, this percentage strongly reduces when it refers to sharing, in favour of other non classified sources. Indeed, in introducing arguments for the discussion, verified users preferably refer to reliable source, while they are less rigorous when sharing others' messages.   
\label{table:verified_global}}}
\end{table}

At a first glance, the majority of the news domains belong to the Reputable category. The second highest percentage is the one of the untagged domains -- UNC. In fact, in our dataset  there are many domains that occur only few times. For example, there are 300 domains that appear in the datasets only once. 

\fasa{\subsubsection{Domain names' analysis of verified users}}

\fasa{While in Section 3 of the Supplementary Material we analysed the domains reputability used in the various verified users communities}, we focus on the urls shared in the  {\it purely political} subcommunities in Table~\ref{table:unv_subcommunity}.
Broadly speaking, we examine the contribution of the different political parties, as represented on Twitter, to the spread of d/misinformation and propaganda.

Table~\ref{table:unv_subcommunity} clearly shows how the vast majority of the news coming from sources considered scarce or non reputable are \fasa{shared} by the center-right and right wing subcommunity (\emph{FI-L-FdI}). Notably, the percentage of non reputable sources shared by the FI-L-FdI accounts is more \fasa{than 30} times the second community in the NR ratio ranking. 

\begin{table}[ht]
\centering
\fasa{
\begin{tabular}{lc|cccc}
Subcommunity & \#url & \text{R} & $\sim$\text{R} & \text{NR} & \text{Others}\\
\hline
\hline
FI-L-FdI & 4759 & 56.4 & 2.3 & 12.8 &  28.5\\
Movimento 5 Stelle & 2385 & 75.5 & 0.1 & 0.4 & 24.0\\
Italia Viva & 857 & 25.3 & 26.6 & 0.1 & 48.0\\
Partito Democratico & 643 & 64.4 & 0.6 & 0.3 & 34.7\\
\hline
\end{tabular}
\smallskip
\caption{\textbf{Domains annotation per political subcommunities -- validated network of verified users.} Frequency of  the various type of news sources. The incidence of Non Reputable sources in the  center-right and right wing parties discursive community is almost 30 times the one of the second community in the NR ratio ranking. In fact, the impact of NR sources is even greater in absolute numbers, due to the greater sharing activity of the users in this group (more than twice the value of the Movimento 5 Stelle subcommunity). For a greater detail in the annotations, consult the Table 4 of the Supplementary Material.}
\label{table:unv_subcommunity}}
\end{table}

\begin{table}[ht]
\centering
\fasa{
\begin{tabular}{lccccc}
Subcommunity & \#post & \#url & \#dist url & \#domain & \#user\\
\hline
\hline
\multicolumn{1}{l}{\texttt{only tweets}} \\ 
FI-L-FdI & 5031 & 4177 & 3728 & 210 & 62 \\
Movimento 5 Stelle & 2406 & 1839 & 1742 & 139 & 103 \\
Italia Viva & 943 & 458 & 417 & 96 & 69 \\
Partito Democratico & 736 & 370 & 353 & 74 & 60 \\
\midrule
\multicolumn{1}{l}{\texttt{only retweets}} \\
FI-L-FdI & 1587 & 582 & 510 & 151 & 72 \\
Movimento 5 Stelle & 997 & 546 & 469 & 104 & 103 \\
Italia Viva & 1048 & 399 & 348 & 147 & 82 \\
Partito Democratico & 747 & 273 & 258 & 94 & 88 \\
\hline
\end{tabular}
\smallskip
\caption{\textbf{Posts, urls, domains and users statistics per political subcommunities -- validated network of verified users:} \#post is the number of posts (divided in tweets and  retweets) by the considered community, \#url is the number of link shared, \#dist url is the number of distinct url, \#domain is the number of distinct domains contained in all urls.  While the number of (validated) verified users is the center-right and right wing subcommunity is lower than any other political group, their activity in writing original posts is at least twice greater than any other group. This difference is not present in the number of retweets.}
\label{table:unv_subcommunity_info}}
\end{table}

Looking at Table~\ref{table:unv_subcommunity_info}, some peculiar behaviours can be still be observed. Again, the center-right and right wing parties, while being the least represented ones in terms of users, are much more active than the other groups: each (verified) user is responsible, on average of almost \fasa{77.86} messages, while the average is 23.96, 22.12 and 15.29 for M5S, IV and PD, respectively. 
It is worth noticing that Italia Viva, while being a recently founded party, is very active; moreover, for them the frequency of quasi reputable sources. 
\fasa{To complete the analysis, in the Supplementary Material the analysis of the hashtag used by the political subcommunity of validated verified users is presented, in order to study the focus of the narratives of the various political groups.}

\subsection{The validated retweet network}
\label{sec:polar-unverif}
\begin{figure}[hb!]
    \centering
\includegraphics[width=.8\linewidth]{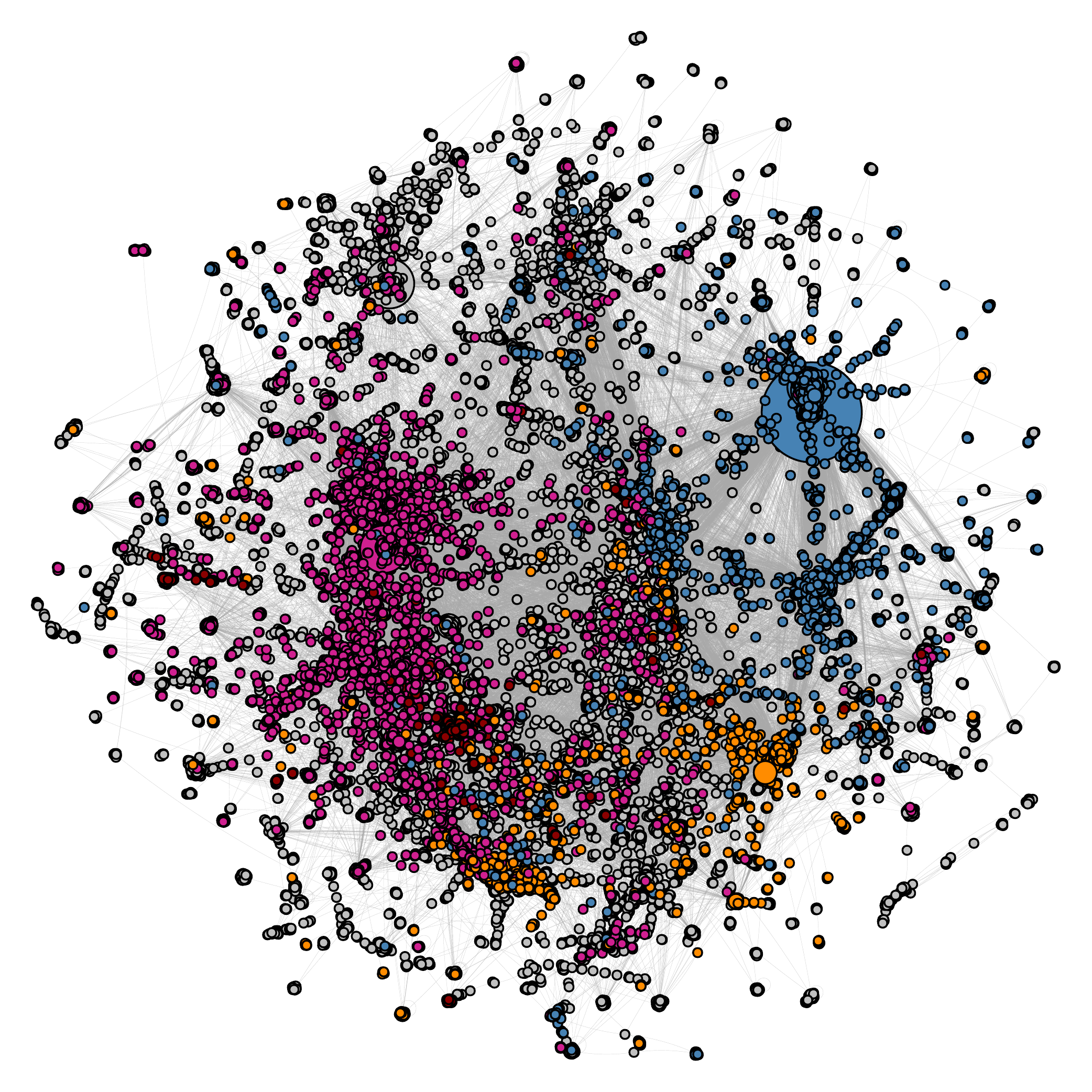}
    \caption{\fasa{\textbf{The directed validated projection of the retweet activity network}: the communities have been highlighted according to their political discursive community they take part to. All nodes not belonging to political discursive communities are in grey. Nodes' dimensions are proportional to their out degree.}\label{fig:dvn}}
\end{figure}
\fasa{In the present subsection we examine the \emph{effective} retweet network, i.e. the activity of user of sharing messages as a reaction to an interesting original tweet. By focusing on the effectiveness of the relations we mean to consider the non random flow of messages from user to user: indeed it may happen that a single tweet is shared simply because it is viral, because its retweeter is particularly active or because the account publishing the original tweet is extremely prolific. Instead we are interested in the flow that cannot be explained only by the activity of users or by the popularity of posts, in order to highlight the non-trivial sharing activity, distinguishing the relevant information from the random noise. 
We define a {\it directed} bipartite network in which one layer is composed by accounts and the other one by the tweets. An arrow connecting a user $u$ to a tweet $t$ represents the $u$ writing the message $t$. The arrow in the opposite direction means that the user $u$ is retweeting the message $t$.
To filter out the random noise from this network, we make use of the directed version of the BiCM, i.e. the Bipartite Directed Configuration Model (\emph{BiDCM}~\cite{DeJeude2018}), described in subsection~\ref{ssec:bidcm}. The BiDCM constrains the in- and out-degree sequences of nodes on both layers, in our representation the users' tweeting and retweeting activity and the virality of posts.
In order to detect the non trivial flow of messages from user to user, for every (directed) couple of accounts, we compared the number of retweets observed in the real system with the expectation of the null model. If the amount of retweets cannot be explained by the theoretical model, we project a link from the author to the retweeter in the monopartite directed network of users. Due to the process of validation, we call this \emph{network directed validated projection}. More details can be found in the subsection~\ref{ssec:validation}.\\ 

In order to infer the affiliation of unverified users to the various discursive communities, we use the  labels obtained in the previous subsection for verified users and propagate them on the validated retweet network using the algorithm proposed in~\cite{Raghavan2007b}. In the Section 6 of the Supplementary Material we show that propagating labels on the entire weighted retweet network, on its binary version or on the validated version is almost equivalent in order to get the labels for the users in the directed validated network.}

After label propagation, the representation of the political communities in the validated retweet network \fasa{(displayed in Fig.~\ref{fig:dvn})} changes  dramatically with respect to the case of the network of verified users: the center-right and right wing community is the most represented community in the whole network, with 11063 users (representing 21.1\% of all the users in the validated network), followed by Italia Viva users with 8035 accounts (15.4\% of all the accounts in the validated network). The impact of M5S and PD is much more limited, with, respectively, 3286 and 564 accounts. It is worth noting that this  result is unexpected, due to the recent formation of Italia Viva. 

As in our previous study targeting the online propaganda~\cite{Caldarelli2020a}, we observe that the most effective users in term of hub score~\cite{Kleinberg1999} are almost exclusively from the center-right and right wing party: considering the first 100 hubs, only 4 are not from this group. Interestingly, 3 out of these 4 are verified users: Roberto Burioni, one of the most famous Italian virologists,  ranking 32nd, Agenzia Ansa, a popular Italian news agency,  ranking 61st, and Tgcom24, the popular newscast of a private TV channel, ranking 73rd. The fourth account is an online news website, ranking 88th: this is a unverified account which belongs to a non political community.

Remarkably, in the top 5 hubs we find 3 of the top 5 hubs already found when considering the online debate on  migrations from northern Africa to Italy~\cite{Caldarelli2020a}: in particular, a journalist of a neo-fascist online newspaper (non verified user), an extreme right activist (non verified user) and the leader of Fratelli d'Italia Giorgia Meloni (verified user), who ranks 3rd in the hub score. Matteo Salvini (verified user), who was the first hub in~\cite{Caldarelli2020a}, ranks 9th, surpassed by his party partner Claudio Borghi (verified user), ranking 6th. The first hub in the present network is an (unverified) extreme right activist, posting videos against African migrants to Italy and accusing them to be responsible of the contagion and of violating lockdown measures.

\subsubsection{Domain analysis on the directed validated network}
Table~\ref{table:dvn_global} shows the annotation results of all the domains tweeted and retweeted by users in the directed validated network. \mape{The annotation was made considering the domains occurring at least 100 times. Even in this case, for those sites not yet evaluated by Newsguard, these have been annotated by the same three members of our team.  We have 100 domains annotated by Newsguard and 53 domains annotated by the three annotators. Also in this case, the annotators showed a moderate agreement for the classification of domains, with $\kappa = 0.57$.}\\ \fasa{There are important differences from Table~\ref{table:verified_global}:} the majority of urls traceable to news sources is \fasa{still} considered reputable\fasa{, but its incidence is much reduced. Interestingly enough, the impact of at least nearly reputable sources is nearly 19\% for original messages and 16\% for retweets, against percentages around 3\% and 2\%, respectively.}\\  

\begin{table}[ht!]
\centering
\fasa{
\begin{tabular}{cc|cccc}
type & \#url & \text{R} & $\sim$\text{R} & \text{NR} & \text{Others}\\
\hline
\hline
tw & 396416 & 39.1 & 11.4 & 7.8 &  41.7\\
rt & 882467 & 51.2 & 7.1 & 8.9 & 32,9 \\
\hline
\end{tabular}
\smallskip
\caption{\textbf{Annotation results over all the domains -- directed validated network}. Differently from the analogous Table~\ref{table:verified_global}, in the case of all users the number of original messages is less than one half of the one of retweets. This behaviour can be explain by the different role that verified users play in the debate: indeed, those accounts are drivers for the discussion and contribute mostly in proposing original messages. Interestingly enough, in passing to considering all users, the percentages of at least nearly reputable sources rise from nearly 3\% and 2\% to nearly 19\% and 16\% for tweet and retweets.\label{table:dvn_global}}}
\end{table}

\begin{table}[ht]
\centering
\fasa{
\begin{tabular}{lcccccc}
Community & \#post & \#url & \#dist url & \#domain & \#user & \#verif\\
\hline
\hline
\multicolumn{1}{l}{\texttt{only tweets}} \\ 
FI-L-FdI  & 176137 & 95902 & 63710 & 3272 & 6831 & 56 \\
Italia Viva & 82356 & 33648 & 25364 & 2243 & 4976 & 56 \\
Movimento 5 Stelle & 41838 & 22940 & 17747 & 1536 & 1974 & 92 \\
Partito Democratico & 3247 & 1759 & 1671 & 277 & 337 & 51 \\
\midrule
\multicolumn{1}{l}{\texttt{only retweets}} \\
FI-L-FdI  & 959748 & 361844 & 54768 & 4304 & 10749 & 48 \\
Italia Viva & 379096 & 121477 & 37084 & 3915 & 7827 & 52 \\
Movimento 5 Stelle & 208195 & 97304 & 27692 & 2647 & 3135 & 72 \\
Partito Democratico & 11517 & 4424 & 3079 & 683 & 528 & 44 \\
\hline
\end{tabular}
\smallskip
\caption{\textbf{Posts, urls, domains and users statistics per political subcommunities  – directed validated network.} Differently from the case on verified users only, in the case of all users the number of tweets is nearly one fifth of the number of retweets.
\label{table:dvn_subcommunity_info}}}
\end{table}

\fasa{Table~\ref{table:dvn_subcommunity_info} reports statistics about  posts, urls, distinct urls, users and verified users in the various political subcommunities in the directed validated network. 
Noticeably, by comparing these numbers with those of Table~\ref{table:unv_subcommunity_info}, reporting analogous statistics about the validated network of verified users, we can see that here the number of retweets is much higher than the one of only tweets, while it was the opposite for verified users: verified users tend to tweet more than retweet, while users in the directed validated network, which comprehends also non verified users, have a greater number of retweets, being even more than ~5 times the one of tweets, depending on the community. It is a behaviour that was already observed in~\cite{Becatti2019,Caldarelli2020a} and it is essentially due to the preeminence of verified users in shaping the public debate on Twitter. It is also remarkable the fact that verified users represent a minority of all users in the verified networks.}

\fasa{Fig.~\ref{fig:overtime-all} shows the trend of the number of posts containing urls over the period of data collection. The highest peak appears after the discovery of the first cases in Lombardy, corresponds to more than 68000 posts containing urls, but a higher traffic is still present before the beginning of the Italian lockdown, while a settling down is present as the quarantine went on\footnote{The low peaks for February 27 and March 10 are due to an interruption in the data collection, caused by a connection breakdown.}. Interestingly, similar trends are present even in the analysis~\cite{Gallotti2020,Chen_2020}.\\ 
It is interesting to note that the incidence of NR sources is nearly constant in the entire period.} 

\begin{figure}[ht!]
    \begin{center}
\includegraphics[width=.8\textwidth]{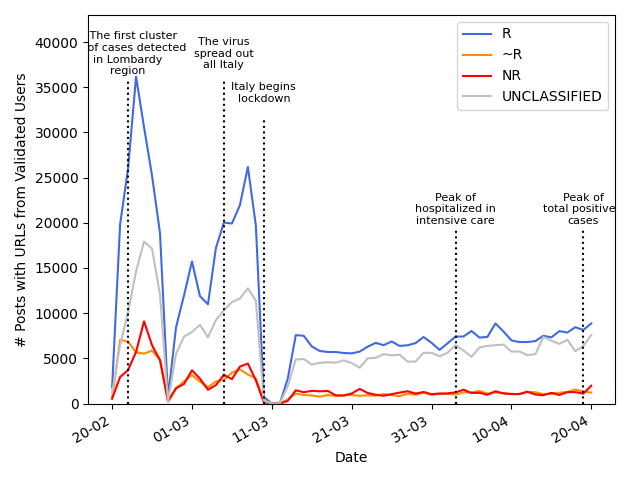}
\end{center}
\fasa{\caption{\textbf{Domains' spreading over time -- validated directed network} The various main event regarding the pandemic have been reported in the plot. It is interesting to notice that the incidence of NR sources in the entire period is more or less constant in time. Interestingly enough, the same reduction of the overall activity after the beginning of the lockdown was detected even in~\cite{Gallotti2020,Chen_2020}.
\label{fig:overtime-all}}}
\end{figure}

 \fasa{Tables~\ref{table:dvn_subcommunity} and~\ref{table:dvn_nr_contribute}  show the core of our analysis, that is, the distribution of reputable and non reputable news sources in the direct validated network, consisting of both verified and non-verified users. Again, we focus directly on the 4 political subcommunities identified in the previous subsection. The incidence of non reputable source in the subcommunity of center-right and right wing parties reach the impressive percentage of 22.1\%, which is even greater than what observed in Table~\ref{table:unv_subcommunity} (i.e. 12.8\%): the contribution of unverified users seems to boost more the diffusion of unreliable contents. It is even more alarming that the percentage of nearly reputable source is great too: considering both non reputable and nearly reputable sources the percentage is of 34.2\%. Thus, more than one third of the urls shared in the validated network by FI-L-FdI subcommunity are at least nearly reputable.}
 
\begin{table}[ht!]
\centering
\fasa{
\begin{tabular}{lc|cccc}
Subcommunity & \#url & \text{R} & $\sim$\text{R} & \text{NR} & \text{Others}\\
\hline
\hline
FI-L-FdI & 457746 & 38.3 & 12.1 & 22.1 & 27.5 \\
Italia Viva & 155125 & 58.7 & 6.7 & 0.7 &  33.9\\
Movimento 5 Stelle & 120244 & 63.8 & 1.4 & 3.1 &31,7\\
Partito Democratico & 6183 & 47.5 & 1.5 & 0.4 &  50.6\\
\hline
\end{tabular}
\smallskip
\caption{\textbf{Domains annotation per political subcommunities – directed validated network} The contribution of urls of the FI-L-FdI community to the validated network is more than 3 times greater than any other discursive community, with 22\% of non reputable urls.  The interested reader can find more details in Table 5 of the Supplementary Material. \label{table:dvn_subcommunity}}}
\end{table}

\begin{table}[h!]
\centering
\begin{tabular}{l|ccccc}
Subcommunity & \#post & \#url & \#domain & \#distinct url & \#user \\
\hline
\hline
\multicolumn{6}{l}{\texttt{only tweets}} \\ 
FI-L-FdI& 29278 & 29324 & 20 & 20059 & 909 \\
Italia Viva & 377 & 378 & 16 & 354 & 84 \\
Movimento 5 Stelle & 608 & 608 & 13 & 448 & 113 \\
Partito Democratico & 3 & 3 & 3 & 3 & 3 \\
\midrule
\multicolumn{1}{l}{\texttt{only retweets}} \\ 
FI-L-FdI & 71613 & 71892 & 20 & 4611 & 6191 \\
Italia Viva & 719 & 719 & 20 & 485 & 453 \\
Movimento 5 Stelle & 3102 & 3105 & 19 & 988 & 736 \\
Partito Democratico & 24 & 24 & 6 & 23 & 13 \\
\hline
\end{tabular}
\smallskip
\fasa{\caption{\textbf{Share of dissemination of NR domains for each of the 4 subcommunities – directed validated network.} Due to its outstanding activity, the contribution of center-right and right wing discursive community to the significant spread of non reputable news sources is impressive: nearly 96\% of all NR urls spread by political subcommunities comes from this group.}
\label{table:dvn_nr_contribute}}
\end{table}

Table~\ref{table:dvn_nr_contribute} offers another point of view. The \#user column shows the number of users of the 4 different political subcommunities who share urls labelled as NR. In absolute numbers, the FI-L-FdI community shares the highest number of NR urls, \fasa{being responsible of the 96\% of NR urls shared by political subcommunities. 
This behaviour is not only due to the the greater amount of users: in the FI-L-FdI subcommunity the accounts sharing NR urls are particularly active. In this group, }the average number of (original) NR posts sent per user is 32.21, which is almost 6 times the average for the M5S users (which has 5.38 NR posts per users); IV  and PD have 4.48 and 1.00 as average, respectively. The frequency of accounts retweeting NR sources among all users from the same community is extremely high too for FI-L-FdI (57.6\% for FI-L-FdI, 23.5\% for M5S, 5.79\% for IV  and 2.5\% for PD -  percentages for the \texttt{only tweets} activity being similar). It thus appears that FI-L-FdI contributes \fasa{substantially} to the diffusion of d/misinformation, not only in relation to the numbers of posts and users, but also in absolute numbers: out of the over 1M tweets, more than 320k tweets refer to a NR url.

\fasa{\subsection{Non Reputable sources shared in the effective flow of misinformation}}
As a final task, over the whole set of tweets produced or shared by the users in the directed validated network, we counted the number of times a message containing a url was shared by users belonging to different political communities, although without considering the semantics of the tweets. Namely, we ignored whether the urls were shared to support or to oppose the presented arguments.

Table~\ref{tab:domains-names-trt} shows the most frequent (tweeted and retweeted) NR domains shared by the political communities; the number of occurrences is reported next to each domain. 

The first NR domains for FI-L-FdI in Table~\ref{tab:domains-names-trt} are related to the right, extreme right and neo-fascist propaganda, as it is the case of  \href{http://www.imolaoggi.it/}{imolaoggi.it}, \href{https://www.ilprimatonazionale.it/}{ilprimatonazionale.it} and \href{http://voxnews.info/}{voxnews.info}, recognised as disinformation websites by NewsGuard  and by the two main Italian debunker websites, \href{https://www.bufale.net/the-black-list-la-lista-nera-del-web/}{bufale.net} and \href{https://www.butac.it/the-black-list/}{BUTAC.it}.

As shown in the table, some domains, although in different number of occurrences, are present under more than one column, thus shared by users close to different political communities. However, since the semantics of the posts in which these domains are present were not investigated, the retweets of the links by more than one political community could be due to contrast, and not to support, the opinions present in the original posts: \fasa{indeed, here we intend to just present the most frequent NR domains}.

\begin{table}[hbt!]
	\scriptsize
	\centering
	\fasa{
	\begin{tabular}{ll|ll|ll|ll}
		\toprule
		\textbf{FI-L-FdI} && \textbf{Italia Viva} && \textbf{Movimento 5 Stelle} && \textbf{Partito Democratico} & \\
		\midrule 
imolaoggi.it & 16041 & dagospia.com & 315 & lantidiplomatico.it & 1114 & it.sputniknews.com & 2\\
ilprimatonazionale.it & 15383 & m.dagospia.com & 134 & m.dagospia.com & 286 & dagospia.com & 2\\
voxnews.info & 9334 & imolaoggi.it & 109 & dagospia.com & 266 & laverita.info & 1\\
stopcensura.info & 8460 & lantidiplomatico.it & 72 & it.sputniknews.com & 98 & lantidiplomatico.it & 1\\
laverita.info & 2647 & ilprimatonazionale.it & 61 & imolaoggi.it & 89 & m.dagospia.com & 1\\
stopcensura.org & 2407 & it.sputniknews.com & 44 & ilprimatonazionale.it & 87 & - & -\\
m.dagospia.com & 2125 & stopcensura.info & 28 & stopcensura.info & 65 & - & -\\
scenarieconomici.it & 1647 & agenpress.it & 25 & voxnews.info & 46 & - & -\\
it.sputniknews.com & 1313 & voxnews.info & 25 & agenpress.it & 37 & - & -\\
dagospia.com & 1291 & laverita.info & 19 & stopcensura.org & 21 & - & -\\
lantidiplomatico.it & 1245 & scenarieconomici.it & 13 & laverita.info & 10 & - & -\\
agenpress.it & 1121 & stopcensura.org & 8 & scenarieconomici.it & 7 & - & -\\
lavocedelpatriota.it & 986 & lavocedelpatriota.it & 6 & lavocedelpatriota.it & 2 & - & -\\
		\toprule
	\end{tabular}
	\caption{\textbf{List of the most frequent NR domains, with relative occurrences, per political subcommunities.} The count was made considering all posts for users of the direct validated network. 
	\label{tab:domains-names-trt}}}
\end{table}

\section{Discussion}
\label{sec:conc}
\fasa{Due to its impact on several dimensions of the society, the online debate regarding the COVID-19 epidemics was the target of several early studies~\cite{Rovetta2020, Celestini2020,Gallotti2020,Cinelli2020,hossain-etal-2020-covidlies,Zhou_2020,Chen_2020,pierri2021vaccinitaly,sharma2020identifying, yang2020covid19}. In the present article we examine the presence of d/misinformation campaigns in the Italian online societal debate about the pandemic during its peak of the first wave. Our analysis is based on a general methodology reviewed in~\cite{Squartinia,Cimini2018} in order to extract both the discursive communities and the effective flow of messages~\cite{Becatti2019,Caldarelli2020a}: in particular, in order to extract both the aforementioned information, we build an entropy-based null-model, constraining part of the information of the real system, and compare the observations on the real network with this benchmark.\\
In particular, we extracted the various discursive communities, focusing our attention on verified users, i.e. public figures whose identity has been checked directly by Twitter platform. Indeed, we observed that, as in other cases~\cite{Becatti2019,Caldarelli2020a,Radicioni2020}, verified accounts lead the debate: their original posts (the \emph{tweets}) are much more than their \emph{retweets}, i.e. their messages sharing others' original tweets. Due to their role in the online debate, we examined in details the activity of verified users. Furthermore, we focused on the \emph{effective} flow of information in the online debate: by comparing the system with an entropy-based null model, we filter out all the random noise associated to the online activities of users.  In this sense, we highlighted all the non trivial retweeting activities and further examined the properties of the filtered network, focusing on the incidence of non reputable sources shared in the debate.}\\

Despite the fact that the results were achieved for a specific country, we believe that \fasa{
our approach, being general and unbiased by construction, is extremely useful to highlight non trivial properties and peculiarities}. In particular, when analyzing the outcome of our investigation,  some features attracted our attention:
\begin{enumerate}
\item {\it Persistence of clusters w.r.t.  different discussion topics:} In Caldarelli et al.~\cite{Caldarelli2020a}, we focused on tweets concerned with immigration, an issue that has been central in the Italian political debate for years. \fasa{In particular, using the same techniques we implemented here in order to extract the effective retweet network, we highlight the presence of coordinated automated accounts increasing \emph{effectively} the visibility of some accounts belonging to the same discursive community.}
Here, we discovered that the clusters and the echo chambers that \fasa{were} detected when analysing tweets about immigration are almost the same as those singled out when considering discussions concerned with  Covid-19\footnote{\fasa{Actually, in the analysis in~\cite{Caldarelli2020a} the center-right and right wing parties were distinct from Media community. In the present analysis they are divided in two different groups only when the first community is further examined by running again a community detection.}}. This may seem surprising, because a discussion about Covid-19 may not be exclusively political, but also medical, social, economic, etc.. From this we can argue that the clusters are political in nature and, even when the topic of discussion changes, users remain in their cluster on Twitter. (It is, in fact, well known that journalists and politicians use Twitter for information and political propaganda, respectively).

The reasons political polarisation and political vision of the world affect so strongly also the analysis of what should be an objective phenomenon is still an intriguing question.
\item {\it (Dis)Similarities amongst offline and online behaviors of members and voters of parties:} Maybe less surprisingly, the political habits is also reflected in the degree of participation to the online discussions.  In particular, among the parties in the center-left wing side, a small party (Italia Viva) shows a much more effective social presence than the larger party of the Italian center-left wing (Partito Democratico), which has many more active members and more parliamentary representation. More generally, there is a significant difference in social presence among the different political parties, and the amount of activity is not at all proportional to the size of the parties in terms of members and voters. 
\item {\it Spread of non reputable news sources:} In the online debate about Covid-19, many links to non reputable (defined such by NewsGuard, a toolkit ranking news website based on criteria of transparency and credibility, led by veteran journalists and news entrepreneurs, \url{https://www.newsguardtech.com/}) news sources are posted and shared. Kind and occurrences of the urls vary with respect to the corresponding political community. Furthermore, \fasa{the center-right and right wing discursive community} is characterised by a relatively small number of verified users that corresponds to a very large number of acolytes which are (on their turn) very active, three times as much as the ones of the opposite communities in the partition. In particular, when considering the amount of retweets from poorly reputable news sites, \fasa{this community} is by far (one order of magnitude) much more active than the others. As noted already in our previous publication~\cite{Caldarelli2020a}, this extra activity could be explained by a more skilled use of the systems of propaganda -- in that case a massive use of bot accounts and a targeted activity against migrants (as resulted from the analysis of the hub list).
\end{enumerate}

\fasa{While our work contributes to the literature regarding the analysis of the impact of d/misinformation on the online societal debate, it paves the path to other crucial analyses. In particular, it is of interest to analyse the structure of the retweet network and how it may contribute to increase the visibility of some of the influential accounts that we detected (this was, in part, the target of the analysis in~\cite{Artime2020}). In this sense, even the role of automatic accounts into the network and in the diffusion of NR source is of utmost importance in order to tackle the problem of online d/misinformation.}

\section{Methods}
\label{sec:bg}

\fasa{In the present section we remind the main steps for the definition of an entropy based null model; the interested reader can refer to the review~\cite{Cimini2018}. We start by revising the Bipartite Configuration Model~\cite{Saracco2015a}, that has been used for detecting the network of similarities of verified users. We are then going to examine the extension of this model to bipartite \emph{directed} networks~\cite{DeJeude2018}. Finally, we present the general methodology to project the information contained in a -directed or undirected- bipartite network, as developed in~\cite{Saracco2016}.}

\fasa{
\subsection{Bipartite Configuration Model}\label{ssec:bicm}
Let us consider a bipartite network $\mathbf{G}^*_\text{Bi}$, in which the two layers are $L$ and $\Gamma$. Define $\mathcal{G}_\text{Bi}$ the ensemble of all possible graphs with the same number of nodes per layer as in $\mathbf{G}^*_\text{Bi}$. It is possible to define the entropy related to the  ensemble as~\citep{park2004statistical}:
\begin{equation}\label{eq:S_undirect}
S=-\sum_{\textbf{G}_\text{Bi}\in\mathcal{G}_\text{Bi}}P(\textbf{G}_\text{Bi})\ln P(\textbf{G}_\text{Bi}), 
\end{equation}
where $P(\textbf{G}_\text{Bi})$ is the probability associated to the instance $\textbf{G}_\text{Bi}$. Now we want to obtain the maximum entropy configuration, constraining some relevant topological information regarding the system. For the bipartite representation of verified and unverified user, a crucial ingredient is the degree sequence, since it is a proxy of the number of interactions (i.e. tweets and retweets) with the other class of accounts. Thus in the present manuscript we focus on the degree sequence. Let us then maximise the entropy  \eqref{eq:S_undirect}, constraining the average over the ensemble of the degree sequence. It can be shown, \cite{Saracco2016}, that the probability distribution over the ensemble is 
\begin{equation}
P(\mathbf{G}_\text{Bi}) = \prod_{i,\alpha} \left(p_{i\alpha}\right)^{m_{i\alpha}}\left(1-p_{i\alpha}\right)^{1-m_{i\alpha}},
\end{equation}
where $m_{i\alpha}$ represent the entries of the biadjacency matrix describing the bipartite network under consideration and $p_{i\alpha}$ is the probability of observing a link between the nodes $i\in L$ and $\alpha\in\Gamma$. The probability $p_{i\alpha}$ can be expressed in terms of the Lagrangian multipliers $x$ and $y$ for nodes on $L$ and $\Gamma$ layers, respectively, as
\begin{equation}\label{app_eq:exact_bicm}
p_{i\alpha}=\dfrac{x_i \, y_\alpha}{1 + x_i \, y_\alpha}.
\end{equation}

In order to obtain the values of $x$ and $y$ that maximize the likelihood to observe the real network, we need to impose the following conditions~\cite{Garlaschelli2008,squartini2011analytical}
\begin{equation}\label{eq:likelihood}
\left\{
\begin{split}
\left\langle k_i \right\rangle &= \sum_{\alpha \in \Gamma} p_{i\alpha} = k_i^* \quad  \forall i \in L\\
\left\langle k_\alpha \right\rangle &= \sum_{i \in L} p_{i\alpha} = k_\alpha^* \quad \forall  \alpha \in \Gamma.\\
\end{split}
\right.,
\end{equation}
where the $*$ indicates quantities measured on the real network.

Actually, the real network is sparse: the bipartite network of verified and unverified users has a connectance $\rho\simeq3.58\times10^{-3}$. In this case the formula (\ref{app_eq:exact_bicm}) can be safely approximated with the Chung-Lu configuration model, i.e. 
\begin{equation*}
    p_{i\alpha}\simeq x_iy_\alpha=\dfrac{k_i^*k_\alpha^*}{m},
\end{equation*}
where $m$ is the total number of links in the bipartite network.

\subsection{Bipartite Directed Configuration Model}
\label{ssec:bidcm}
In the present subsection we will consider the case of the extension of the BiCM to \emph{direct} bipartite networks and highlight the peculiarities of the network under analysis in this representation. The adjancency matrix describing a direct bipartite network of layers $L$ and $\Gamma$ has a peculiar block structure, once nodes are order by layer membership (here the nodes on $L$ layer first):
\begin{equation}
\textbf{A}=
\left(
\begin{array}{c|c}
\textbf{O} & \textbf{M}\\

\hline

\textbf{N}^\text{T}& \textbf{O}
\end{array}
\right),
\end{equation}
where the $\textbf{O}$ blocks represent null matrices (indeed they describe links connecting nodes inside the same layer: by construction they are exactly zero) and $\textbf{M}$ and $\textbf{N}$ are non zero blocks, describing links connecting nodes on layer $L$ with those on layer $\Gamma$ and viceversa. In general $\textbf{M}\neq\textbf{N}$, otherwise the network is not distinguishable from an undirected one.

We can perform the same machinery of the section above, but for the extension of the degree sequence to a directed degree sequence, i.e. considering the in- and out-degrees for nodes on the layer $L$, 
\begin{equation}
    k_i^\text{out}=\sum_{\alpha \in \Gamma}m_{i\alpha}\quad\text{and}\quad k_i^\text{in}=\sum_{\alpha \in \Gamma}n_{i\alpha}
\end{equation}
(here $m_{i\alpha}$ and $n_{i\alpha}$ represent respectively the entry of matrices $\mathbf{M}$ and $\mathbf{N}$) and for nodes on the layer $\Gamma$, 
\begin{equation}
    k_\alpha^\text{out}=\sum_{i \in L}n_{i\alpha}\quad\text{and}\quad k_\alpha^\text{in}=\sum_{i \in L}m_{i\alpha}.
\end{equation}

The definition of the Bipartite \emph{Directed} Configuration Model (BiDCM,~\cite{DeJeude2018}), i.e. the extension of the BiCM above, follows closely the same steps described in the previous subsection. Interestingly enough, the probabilities relative to the presence of links from $L$ to $\Gamma$ are independent on the probabilities relative to the presence of links from $\Gamma$ to $L$.
If $q_{i\alpha}$ is the probability of observing a link from node $i$ to node $\alpha$ and $q'_{i\alpha}$ the probability of observing a link in the opposite direction, we have
\begin{equation}
    q_{i\alpha}=\dfrac{x_i^\text{out}y_\alpha^\text{in}}{1+x_i^\text{out}y_\alpha^\text{in}}\quad\text{and}\quad q'_{i\alpha}=\dfrac{x_i^\text{in}y_\alpha^\text{out}}{1+x_i^\text{in}y_\alpha^\text{out}}, 
\end{equation}
where $x_i^\text{out}$ and $x_i^\text{in}$ are the Lagrangian multipliers relative to the node $i\in L$, respectively for the out- and the in-degrees, and $y_\alpha^\text{out}$ and $y_\alpha^\text{in}$ are the analogous for $\alpha\in\Gamma$.

In the present application we have some simplifications: the bipartite directed network representation describes users (on one layer) writing and retweeting posts (on the other layer). If users are on the layer $L$ and posts on the opposite one and $m_{i\alpha}$ represents the user $i$ writing the post $\alpha$, then $k_\alpha^\text{in}=1\,\forall \alpha\in\Gamma$, since each message cannot have more than an author. Notice that, since our constraints are conserved on average, we are considering, in the ensemble of all possible realisations, even instances in which $k_\alpha^\text{in}>1$ or $k_\alpha^\text{in}=0$, or, otherwise stated, non physical; nevertheless the average is constrained to the right value, i.e. 1. The fact that $k_\alpha^\text{in}$ is the same for every $\alpha$ allows for a great simplification of the probability per link on $\mathbf{M}$:
\begin{equation}\label{eq:culo}
q_{i\alpha}=\dfrac{(k_i^\text{out})^*}{N_\Gamma},
\end{equation}
where $N_\Gamma$ is the total number of nodes on the $\Gamma$ layer. The simplification in (\ref{eq:culo}) is extremely helpful in the projected validation of the bipartite directed network~\cite{Becatti2019}.

\subsection{Validation of the projected network}
\label{ssec:validation}
The information contained in a bipartite -directed or undirected- network, can be projected onto one of the two layers. The rationale is to obtain a monopartite network encoding the non trivial interactions among the two layers of the original bipartite network.  
The method is pretty general, once we have a null model in which probabilities per link are independent, as it is the case of both BiCM and BiDCM~\cite{Saracco2016}. The method is graphically depicted in Fig.~\ref{fig:user_polarization_network} in the case of BiCM; the case of BiDCM is analogous.

The first step is represented by the definition of a bipartite motif that may capture the non trivial similarity (in the case of an undirected bipartite network) or flux of information (in the case of a directed bipartite network). This quantity can be captured by the number of $V-$motifs between users $i$ and $j$~\cite{Diestel2006, Saracco2015a},
\begin{equation}
    V_{ij} = \sum_{\alpha \in \Gamma} m_{i\alpha}m_{j\alpha}, 
\end{equation}
or by its direct extension 
\begin{equation}
    \mathcal{V}_{ij}=\sum_{\alpha\in \Gamma}m_{i\alpha}n_{\alpha j}
\end{equation}
(note that $\mathcal{V}_{ij}\neq\mathcal{V}_{ji}$). We compare the abundance of these motifs with the null models defined above: all motifs that cannot be explained by the null model, i.e. whose p-value are statistically significance, are validated into the projection on one of the layers~\cite{Saracco2016}. 

In order to assess the statistically significance of the observed motifs, we calculate the distribution associated to the various motifs. For instance, the expected value for the number of V-motifs connecting $i$ and $j$ in an undirected bipartite network is
\begin{equation}
    \left\langle V_{ij} \right\rangle =  \sum_{\alpha \in \Gamma} p_{i\alpha} \, p_{j\alpha},
\end{equation}
where $p_{i\alpha}$s are the probability of the BiCM. Analogously, 
\begin{equation}\label{eq:exp_vmotif}
\left\langle \mathcal{V}_{ij} \right\rangle = \sum_{p \in P} q_{i\alpha} \, q'_{j\alpha}=\dfrac{(k_i^\text{out})^* (k_j^\text{in})^*}{N_\Gamma},
\end{equation}
where in the last step we use the simplification of (\ref{eq:culo})~\cite{Becatti2019}.

In both the direct and the undirect case, the distribution of the V-motifs or of the directed extensions is Poisson Binomial one, i.e. a binomial distribution in which each event shows a different probability. In the present case, due to the sparsity of the analysed networks, we can safely approximate the Poisson-Binomial distribution with a Poisson one~\cite{Hong2013}.\\ 

In order to state the statistical significance of the observed value, we calculate the related p-values according to the relative null-models. Once we have a p-value for every detected V-motif, the related statistical significance can be established through the False Discovery Rate (\emph{FDR}) procedure~\cite{benjamini1995controlling}, which, respect to other multiple test hypothesis, controls the number of False Positives. In our case, all rejected hypotheses identify the amount of V-motifs that cannot be explained only by the ingredients of the null model and thus carry non trivial information regarding the systems. In this sense, the validated projected network includes a link for every rejected hypothesis, connecting the nodes involved in the related motifs. 
}

\begin{figure}[ht!]
\begin{tikzpicture}[>=stealth,shorten >=2pt,auto,node distance=3cm,main node/.style={circle,draw},scale=0.85]
\node at (1.75, 1.5) {Real Network};

\node[main node,fill=orange,color=orange] (1) at (.5, -1.5) {};
\node[main node,fill=orange,color=orange] (2) at (1.5, -1.5) {}; 
\node[main node,fill=orange,color=orange] (3) at (2.5, -1.5) {}; 
\node[main node,fill=orange,color=orange] (4) at (3.5, -1.5) {}; 

\node[main node,fill=teal,color=teal] (5) at (1, 0) {}; 
\node[main node,fill=teal,color=teal] (6) at (2, 0) {}; 
\node[main node,fill=teal,color=teal] (7) at (3, 0) {}; 

\node at (1, 0.5) {$i$};
\node at (2, 0.5) {$j$};

\node at (0, -0.65) {\textbf{a)}};

\draw[-] (1) edge node {} (5);
\draw[-] (2) edge node {} (5);
\draw[-] (3) edge node {} (5);

\draw[-] (1) edge node {} (6);
\draw[-] (2) edge node {} (6);
\draw[-] (3) edge node {} (6);
\draw[-] (4) edge node {} (6);

\draw[-] (3) edge node {} (7);
\draw[-] (4) edge node {} (7);


\node[main node,fill=teal,color=teal] (12) at (1, -4) {}; 
\node[main node,fill=teal,color=teal] (13) at (2, -4) {}; 
\node[main node,fill=teal,color=teal] (14) at (3, -4) {}; 

\node[circle,fill=teal,draw=purple] (12) at (1, -4) {}; 
\node[circle,fill=teal,draw=purple] (13) at (2, -4) {};

\node[main node,fill=orange,color=orange] (8) at (.5, -5.5) {};
\node[main node,fill=orange,color=orange] (9) at (1.5, -5.5) {}; 
\node[main node,fill=orange,color=orange] (10) at (2.5, -5.5) {}; 
\node[main node,fill=orange,color=orange] (11) at (3.5, -5.5) {};

\node at (1, -3.5) {$i$};
\node at (2, -3.5) {$j$};
\node at (0, -4.65) {\textbf{c)}};

\draw[-] (8) edge node {} (12);
\draw[dashed, color=purple, line width=1] (8) edge node {} (12);
\draw[-] (9) edge node {} (12);
\draw[dashed, color=purple, line width=1] (9) edge node {} (12);
\draw[-] (10) edge node {} (12);
\draw[dashed, color=purple, line width=1] (10) edge node {} (12);

\draw[-] (8) edge node {} (13);
\draw[dashed, color=purple, line width=1] (8) edge node {} (13);
\draw[-] (9) edge node {} (13);
\draw[dashed, color=purple, line width=1] (9) edge node {} (13);
\draw[-] (10) edge node {} (13);
\draw[dashed, color=purple, line width=1] (10) edge node {} (13);
\draw[-] (11) edge node {} (13);

\draw[-] (10) edge node {} (14);
\draw[-] (11) edge node {} (14);

\draw [rounded corners] (-0.5,2) rectangle (4,-6.5);
\draw [->, rounded corners] (1.75,-6.5) -- (1.75,-8.5) -- (4,-8.5);

\node at (9.25, 1.5) {BiCM};
\draw [rounded corners] (5.5,2) rectangle (13,-6.5);
\draw [->, rounded corners] (9.25,-6.5) -- (9.25,-8.5) -- (8,-8.5);

\node at (9.75, 0) {$\cdots$};
\node at (6, -0.65) {\textbf{b)}};
\draw [fill=teal,color=teal] (7.25,0.25) circle [radius=0.1];
\draw [fill=teal,color=teal] (7.75,0.25) circle [radius=0.1];
\draw [fill=teal,color=teal] (8.25,0.25) circle [radius=0.1];

\draw [fill=orange,color=orange] (7,-0.25) circle [radius=0.1];
\draw [fill=orange,color=orange] (7.5,-0.25) circle [radius=0.1];
\draw [fill=orange,color=orange] (8,-0.25) circle [radius=0.1];
\draw [fill=orange,color=orange] (8.5, -0.25) circle [radius=0.1];

\node at (7.25, 0.75) {$i$};
\node at (7.75, 0.75) {$j$};


\draw(11.25,0.25) -- (11.,-0.25);
\draw (11.75,0.25) -- (12.,-0.25);
\draw (11.75,0.25) -- (11.5,-0.25);
\draw (12.25,0.25) -- (11.5,-0.25);
\draw(11.25,0.25) -- (12.5,-0.25);

\draw [fill=teal,color=teal] (11.25,0.25) circle [radius=0.1];
\draw [fill=teal,color=teal] (11.75,0.25) circle [radius=0.1];
\draw [fill=teal,color=teal] (12.25,0.25) circle [radius=0.1];

\draw [fill=orange,color=orange] (11,-0.25) circle [radius=0.1];
\draw [fill=orange,color=orange] (11.5,-0.25) circle [radius=0.1];
\draw [fill=orange,color=orange] (12,-0.25) circle [radius=0.1];
\draw [fill=orange,color=orange] (12.5, -0.25) circle [radius=0.1];

\node at (11.75, 0.75) {$j$};
\node at (11.25, 0.75) {$i$};

\node at (9.75, -1.75) {$\cdots$};
\node at (9.75, -0.75) {$\ddots$};

\draw(7.25,-1.5) -- (7.,-2);
\draw(7.25,-1.5) -- (8.,-2);
\draw (7.75,-1.5) -- (8,-2);
\draw (7.75,-1.5) -- (7.5,-2);
\draw (8.25,-1.5) -- (7.5,-2);
\draw(8.25,-1.5) -- (8.5,-2);

\draw [fill=teal,color=teal] (7.25,-1.5) circle [radius=0.1];
\draw [fill=teal,color=teal] (7.75,-1.5) circle [radius=0.1];
\draw [fill=teal,color=teal] (8.25,-1.5) circle [radius=0.1];

\draw [fill=orange,color=orange] (7,-2) circle [radius=0.1];
\draw [fill=orange,color=orange] (7.5,-2) circle [radius=0.1];
\draw [fill=orange,color=orange] (8,-2) circle [radius=0.1];
\draw [fill=orange,color=orange] (8.5, -2) circle [radius=0.1];

\node at (7.75, -1) {$j$};
\node at (7.25, -1) {$i$};


\draw(11.25,-1.5) -- (11,-2);
\draw(11.25,-1.5) -- (11.5,-2);
\draw(11.25,-1.5) -- (12,-2);
\draw(11.25,-1.5) -- (12.5,-2);

\draw(11.75,-1.5) -- (11,-2);
\draw(11.75,-1.5) -- (11.5,-2);
\draw(11.75,-1.5) -- (12,-2);
\draw(11.75,-1.5) -- (12.5,-2);

\draw(12.25,-1.5) -- (11,-2);
\draw(12.25,-1.5) -- (11.5,-2);
\draw(12.25,-1.5) -- (12,-2);
\draw(12.25,-1.5) -- (12.5,-2);

\draw [fill=teal,color=teal] (11.25,-1.5) circle [radius=0.1];
\draw [fill=teal,color=teal] (11.75,-1.5) circle [radius=0.1];
\draw [fill=teal,color=teal] (12.25,-1.5) circle [radius=0.1];

\draw [fill=orange,color=orange] (11,-2) circle [radius=0.1];
\draw [fill=orange,color=orange] (11.5,-2) circle [radius=0.1];
\draw [fill=orange,color=orange] (12,-2) circle [radius=0.1];
\draw [fill=orange,color=orange] (12.5, -2) circle [radius=0.1];

\node at (11.75, -1) {$j$};
\node at (11.25, -1) {$i$};

\node at (9.75, -4) {$\cdots$};
\node at (6, -4.65) {\textbf{d)}};

\draw [fill=teal,color=teal] (7.25,-3.75) circle [radius=0.1];
\draw [color=purple] (7.25,-3.75) circle [radius=0.1];
\draw [fill=teal,color=teal] (7.75,-3.75) circle [radius=0.1];
\draw [color=purple] (7.75,-3.75) circle [radius=0.1];
\draw [fill=teal,color=teal] (8.25,-3.75) circle [radius=0.1];

\draw [fill=orange,color=orange] (7,-4.25) circle [radius=0.1];
\draw [fill=orange,color=orange] (7.5,-4.25) circle [radius=0.1];
\draw [fill=orange,color=orange] (8,-4.25) circle [radius=0.1];
\draw [fill=orange,color=orange] (8.5, -4.25) circle [radius=0.1];

\node at (7.75, -3.25) {$j$};
\node at (7.25, -3.25) {$i$};


\draw(11.25,-3.75) -- (11,-4.25);
\draw (11.75,-3.75) -- (12,-4.25);
\draw (11.75,-3.75) -- (11.5,-4.25);
\draw (12.25,-3.75) -- (11.5,-4.25);
\draw(11.25,-3.75) -- (12.5,-4.25);

\draw [fill=teal,color=teal] (11.25,-3.75) circle [radius=0.1];
\draw [color=purple] (11.25,-3.75) circle [radius=0.1];
\draw [fill=teal,color=teal] (11.75,-3.75) circle [radius=0.1];
\draw [color=purple] (11.75,-3.75) circle [radius=0.1];
\draw [fill=teal,color=teal] (12.25,-3.75) circle [radius=0.1];

\draw [fill=orange,color=orange] (11,-4.25) circle [radius=0.1];
\draw [fill=orange,color=orange] (11.5,-4.25) circle [radius=0.1];
\draw [fill=orange,color=orange] (12,-4.25) circle [radius=0.1];
\draw [fill=orange,color=orange] (12.5, -4.25) circle [radius=0.1];

\node at (11.75, -3.25) {$j$};
\node at (11.25, -3.25) {$i$};

\node at (9.75, -5.75) {$\cdots$};
\node at (9.75, -4.75) {$\ddots$};

\draw(7.25,-5.5) -- (7,-6);
\draw(7.25,-5.5) -- (8,-6);

\draw[dashed,color=purple, line width=1](7.25,-5.5) -- (8,-6);

\draw (7.75,-5.5) -- (8,-6);
\draw (7.75,-5.5) -- (7.5,-6);

\draw[dashed,color=purple, line width=1] (7.75,-5.5) -- (8,-6);

\draw (8.25,-5.5) -- (7.5,-6);
\draw(8.25,-5.5) -- (8.5,-6);

\draw [fill=teal,color=teal] (7.25,-5.5) circle [radius=0.1];
\draw [color=purple] (7.25,-5.5) circle [radius=0.1];
\draw [fill=teal,color=teal] (7.75,-5.5) circle [radius=0.1];
\draw [color=purple] (7.75,-5.5) circle [radius=0.1];
\draw [fill=teal,color=teal] (8.25,-5.5) circle [radius=0.1];

\draw [fill=orange,color=orange] (7,-6) circle [radius=0.1];
\draw [fill=orange,color=orange] (7.5,-6) circle [radius=0.1];
\draw [fill=orange,color=orange] (8,-6) circle [radius=0.1];
\draw [fill=orange,color=orange] (8.5, -6) circle [radius=0.1];

\node at (7.75, -5) {$j$};
\node at (7.25, -5) {$i$};


\draw(11.25,-5.5) -- (11,-6);
\draw[dashed,color=purple, line width=1] (11.25,-5.5) -- (11,-6);
\draw(11.25,-5.5) -- (11.5,-6);
\draw[dashed,color=purple, line width=1] (11.25,-5.5) -- (11.5,-6);
\draw(11.25,-5.5) -- (12,-6);
\draw[dashed,color=purple, line width=1] (11.25,-5.5) -- (12,-6);
\draw(11.25,-5.5) -- (12.5,-6);
\draw[dashed,color=purple, line width=1](11.25,-5.5) -- (12.5,-6);

\draw(11.75,-5.5) -- (11,-6);
\draw(11.75,-5.5) -- (11.5,-6);
\draw(11.75,-5.5) -- (12,-6);
\draw(11.75,-5.5) -- (12.5,-6);

\draw[dashed,color=purple, line width=1](11.75,-5.5) -- (11,-6);
\draw[dashed,color=purple, line width=1](11.75,-5.5) -- (11.5,-6);
\draw[dashed,color=purple, line width=1](11.75,-5.5) -- (12,-6);
\draw[dashed,color=purple, line width=1](11.75,-5.5) -- (12.5,-6);

\draw(12.25,-5.5) -- (11,-6);
\draw(12.25,-5.5) -- (11.5,-6);
\draw(12.25,-5.5) -- (12,-6);
\draw(12.25,-5.5) -- (12.5,-6);

\draw [fill=teal,color=teal] (11.25,-5.5) circle [radius=0.1];
\draw [color=purple] (11.25,-5.5) circle [radius=0.1];
\draw [fill=teal,color=teal] (11.75,-5.5) circle [radius=0.1];
\draw [color=purple] (11.75,-5.5) circle [radius=0.1];
\draw [fill=teal,color=teal] (12.25,-5.5) circle [radius=0.1];

\draw [fill=orange,color=orange] (11,-6) circle [radius=0.1];
\draw [fill=orange,color=orange] (11.5,-6) circle [radius=0.1];
\draw [fill=orange,color=orange] (12,-6) circle [radius=0.1];
\draw [fill=orange,color=orange] (12.5, -6) circle [radius=0.1];

\node at (11.75, -5) {$j$};
\node at (11.25, -5) {$i$};

\draw [rounded corners] (4,-7) rectangle (8,-10);

\node at (6,-7.5) {Validated Projection};

\node at (6.5,-8.5) {$j$};
\node at (6.5-0.866,-9+0.866) {$i$};
\node at (4.5,-9.) {\textbf{e)}};

\draw[color=purple, line width=1](6.5-0.866,-9.5+0.866) -- (6.5,-9.);
\draw [fill=teal,color=teal] (6.5,-9.) circle [radius=0.1];
\draw [color=purple] (6.5,-9) circle [radius=0.1];
\draw [fill=teal,color=teal] (6.5-0.866,-9.5+0.866) circle [radius=0.1];
\draw [color=purple] (6.5-0.866,-9.5+0.866) circle [radius=0.1];

\draw [fill=teal,color=teal] (6.5-0.866,-9.5) circle [radius=0.1];
\end{tikzpicture}

\caption{\textbf{Schematic representation of the projection procedure for
bipartite undirected networks}. a) An example of a real bipartite network. For the actual application, the two layers represent verified (turquoise) and unverified (gray) users and a link between nodes of different layers 
is present if one of the two users retweeted the other one, at least once. b) Definition of the Bipartite Configuration Model (BiCM) ensemble. Such ensemble includes all possible link realisations, once the number of nodes per layers has been fixed.
c) we focus our attention on nodes $i$ and $j$, i.e., two verified users, and count the number of common neighbours (in magenta both the nodes and the links to their common neighbours). Subsequently, d) we compare this measure on the real network with the one on the ensemble: If this overlap is statistically significant with respect to the BiCM, e) we have a link connecting the two verified users in the projected network. \fasa{The figure is an adaptation from~\cite{Caldarelli2020a}}.}
\label{fig:user_polarization_network}
\end{figure}

\section{Availability of Data and Materials}
The data that support the findings of this study are available from Twitter, but restrictions apply to the availability of these data, which were used under license for the current study, and so are not publicly available. Data are however available from the authors upon reasonable request and with permission of Twitter. \fasa{NewsGuard data are proprietary and cannot be shared.}
\section{Competing interests}
The authors have no competing interest on this paper
\section{Funding}
GC acknwoledge support from ITALY-ISRAEL project "Mac2Mic" and EU project nr. 952026 - HumanE-AI-Net. \fasa{FS acknowledges support from the EU project SoBigData-PlusPlus, nr.\ 871042.} All the authors acknowledge support from IMT PAI "Toffee". \section{Authors' contributions}
All the authors devise the experiment and wrote the paper, FS, MPe, MPr collected and analysed the data.

\bibliographystyle{spphys}       
\bibliography{biblio}   

\begin{thebibliography}{10}
\providecommand{\url}[1]{{#1}}
\providecommand{\urlprefix}{URL }
\expandafter\ifx\csname urlstyle\endcsname\relax
  \providecommand{\doi}[1]{DOI \discretionary{}{}{}#1}\else
  \providecommand{\doi}{DOI \discretionary{}{}{}\begingroup
  \urlstyle{rm}\Url}\fi

\bibitem{Bradshaw2018b}
S.~Bradshaw, P.~Howard, How does junk news spread so quickly across social
  media? {A}lgorithms, advertising and exposure in public life, Oxford Internet
  Institute -- White Paper  (2018)

\bibitem{Gonzalez-Bailon2013}
S.~Gonz{\'{a}}lez-Bail{\'{o}}n, J.~Borge-Holthoefer, Y.~Moreno, {Broadcasters
  and Hidden Influentials in Online Protest Diffusion}, Am. Behav. Sci.
  \textbf{57}(7), 943 (2013).
\newblock \doi{10.1177/0002764213479371}

\bibitem{cresci2015fame}
S.~Cresci, R.~Di~Pietro, M.~Petrocchi, A.~Spognardi, M.~Tesconi, Fame for sale:
  efficient detection of fake twitter followers, Decision Support Systems
  \textbf{80}, 56 (2015)

\bibitem{Stella2018}
M.~Stella, M.~Cristoforetti, M.~De~Domenico, Influence of augmented humans in
  online interactions during voting events, PLOS ONE \textbf{14}(5), 1 (2019).
\newblock \doi{10.1371/journal.pone.0214210}

\bibitem{Ciampaglia2018}
G.L. Ciampaglia, A.~Nematzadeh, F.~Menczer, A.~Flammini, {How algorithmic
  popularity bias hinders or promotes quality}, Sci. Rep.  (2018).
\newblock \doi{10.1038/s41598-018-34203-2}

\bibitem{Ferrara2016rise}
E.~Ferrara, O.~Varol, C.~Davis, F.~Menczer, A.~Flammini, The rise of social
  bots, Commun. ACM \textbf{59}(7), 96 (2016)

\bibitem{Ferrara2019}
K.~Yang, O.~Varol, C.A. Davis, E.~Ferrara, A.~Flammini, F.~Menczer, Arming the
  public with {AI} to counter social bots, CoRR \textbf{abs/1901.00912} (2019).
\newblock \urlprefix\url{http://arxiv.org/abs/1901.00912}

\bibitem{Cresci2019WebSci}
S.~Cresci, M.~Petrocchi, A.~Spognardi, S.~Tognazzi, in \emph{11th International
  ACM Web Science Conference} (2019), pp. 47--56

\bibitem{Bovet2019}
A.~Bovet, H.A. Makse, {Influence of fake news in Twitter during the 2016 US
  presidential election}, Nat. Commun. \textbf{10}(1) (2019)

\bibitem{Becatti2019}
C.~Becatti, G.~Caldarelli, R.~Lambiotte, F.~Saracco, {Extracting significant
  signal of news consumption from social networks: the case of {T}witter in
  {I}talian political elections}, Palgrave Commun.  (2019)

\bibitem{Caldarelli2020a}
G.~Caldarelli, R.~{De Nicola}, F.~{Del Vigna}, M.~Petrocchi, F.~Saracco, {The
  role of bot squads in the political propaganda on Twitter}, Commun. Phys.
  \textbf{3}(1), 1 (2020).
\newblock \doi{10.1038/s42005-020-0340-4}

\bibitem{Rovetta2020}
A.~Rovetta, A.S. Bhagavathula, {COVID-19-related web search behaviors and
  infodemic attitudes in Italy: Infodemiological study}, J. Med. Internet Res.
  (2020).
\newblock \doi{10.2196/19374}

\bibitem{Celestini2020}
A.~Celestini, M.~{Di Giovanni}, S.~Guarino, F.~Pierri, {Information disorders
  on Italian Facebook during COVID-19 infodemic}, arXiv  (2020).
\newblock \urlprefix\url{http://arxiv.org/abs/2007.11302}

\bibitem{Gallotti2020}
R.~Gallotti, F.~Valle, N.~Castaldo, P.~Sacco, M.~{De Domenico}, {Assessing the
  risks of ‘infodemics' in response to COVID-19 epidemics}, Nat. Hum. Behav.
  \textbf{4}(12), 1285 (2020).
\newblock \doi{10.1038/s41562-020-00994-6}.
\newblock \urlprefix\url{https://doi.org/10.1038/s41562-020-00994-6}

\bibitem{Cinelli2020}
M.~Cinelli, W.~Quattrociocchi, A.~Galeazzi, C.M. Valensise, E.~Brugnoli, A.L.
  Schmidt, P.~Zola, F.~Zollo, A.~Scala, {The COVID-19 social media infodemic},
  Sci. Rep. \textbf{10}(1), 16598 (2020).
\newblock \doi{10.1038/s41598-020-73510-5}.
\newblock \urlprefix\url{www.nature.com/scientificreports}

\bibitem{Radicioni2020}
T.~Radicioni, E.~Pavan, T.~Squartini, F.~Saracco, {Analysing Twitter Semantic
  Networks: the case of 2018 Italian Elections}, arXiv  (2020).
\newblock \urlprefix\url{http://arxiv.org/abs/2009.02960}

\bibitem{Squartinia}
T.~Squartini, D.~Garlaschelli, \emph{{Maximum-entropy networks. Pattern
  detection, network reconstruction and graph combinatorics}} (Springer
  International Publishing, 2017)

\bibitem{Cimini2018}
G.~Cimini, T.~Squartini, F.~Saracco, D.~Garlaschelli, A.~Gabrielli,
  G.~Caldarelli, {The Statistical Physics of Real-World Networks}, Nat. Rev.
  Phys. \textbf{1}(1), 58 (2018).
\newblock \doi{10.1038/s42254-018-0002-6}

\bibitem{echo2016}
S.~Flaxman, S.~Goel, J.M. Rao, {Filter Bubbles, Echo Chambers, and Online News
  Consumption}, Public Opinion Quarterly \textbf{80}(S1), 298 (2016).
\newblock \doi{10.1093/poq/nfw006}

\bibitem{hossain-etal-2020-covidlies}
T.~Hossain, R.L. Logan~IV, A.~Ugarte, Y.~Matsubara, S.~Young, S.~Singh, in
  \emph{Proceedings of the 1st Workshop on {NLP} for {COVID}-19 (Part 2) at
  {EMNLP} 2020} (Association for Computational Linguistics, Online, 2020).
\newblock \doi{10.18653/v1/2020.nlpcovid19-2.11}.
\newblock \urlprefix\url{https://www.aclweb.org/anthology/2020.nlpcovid19-2.11}

\bibitem{Zhou_2020}
X.~Zhou, A.~Mulay, E.~Ferrara, R.~Zafarani, {ReCOVery: A Multimodal Repository
  for COVID-19 News Credibility Research}, Proceedings of the 29th ACM
  International Conference on Information \& Knowledge Management  (2020).
\newblock \doi{10.1145/3340531.3412880}.
\newblock \urlprefix\url{http://dx.doi.org/10.1145/3340531.3412880}

\bibitem{Chen_2020}
E.~Chen, K.~Lerman, E.~Ferrara, {Tracking Social Media Discourse About the
  COVID-19 Pandemic: Development of a Public Coronavirus Twitter Data Set},
  JMIR Public Health Surveill \textbf{6}(2) (2020).
\newblock \doi{10.2196/19273}

\bibitem{pierri2021vaccinitaly}
F.~Pierri, S.~Pavanetto, M.~Brambilla, S.~Ceri.
\newblock Vaccinitaly: monitoring italian conversations around vaccines on
  twitter (2021)

\bibitem{sharma2020identifying}
K.~Sharma, E.~Ferrara, Y.~Liu, Identifying coordinated accounts in
  disinformation campaigns, CoRR \textbf{abs/2008.11308} (2020).
\newblock \urlprefix\url{https://arxiv.org/abs/2008.11308}

\bibitem{yang2020covid19}
K.C. Yang, F.~Pierri, P.M. Hui, D.~Axelrod, C.~Torres-Lugo, J.~Bryden,
  F.~Menczer.
\newblock The covid-19 infodemic: Twitter versus facebook (2020)

\bibitem{AGCOM2017}
AGCOM.
\newblock {Journalism Observatory II edition} (2017).
\newblock
  \urlprefix\url{https://www.agcom.it/documentazione/documento?p{\_}p{\_}auth=fLw7zRht{\&}p{\_}p{\_}id=101{\_}INSTANCE{\_}FnOw5lVOIXoE{\&}p{\_}p{\_}lifecycle=0{\&}p{\_}p{\_}col{\_}id=column-1{\&}p{\_}p{\_}col{\_}count=1{\&}{\_}101{\_}INSTANCE{\_}FnOw5lVOIXoE{\_}struts{\_}action={\%}2Fasset{\_}publisher{\%}2Fview{\_}content{\&}{\_}101{\_}INSTANCE{\_}FnOw5lVOIXoE{\_}asse}

\bibitem{AGCOM2018}
AGCOM, Report on the consumption of information.
\newblock Tech. Rep. February, Autorit\`{a} per le Garanzie delle Comunicazioni
  (2018)

\bibitem{Adamic2005}
L.A. Adamic, N.S. Glance, in \emph{3rd International Workshop on Link
  discovery, LinkKDD 2005, Chicago, Illinois, USA, August 21-25, 2005} (2005),
  pp. 36--43

\bibitem{Conover2011}
M.~Conover, J.~Ratkiewicz, M.~Francisco, {Political polarization on twitter.},
  Icwsm  (2011).
\newblock \doi{10.1021/ja202932e}

\bibitem{Conover2011a}
M.D. Conover, B.~Gon{\c{c}}alves, J.~Ratkiewicz, A.~Flammini, F.~Menczer, in
  \emph{Proc. - 2011 IEEE Int. Conf. Privacy, Secur. Risk Trust IEEE Int. Conf.
  Soc. Comput. PASSAT/SocialCom 2011} (2011).
\newblock \doi{10.1109/PASSAT/SocialCom.2011.34}

\bibitem{Conover2012}
M.D. Conover, B.~Gon{\c{c}}alves, A.~Flammini, F.~Menczer, {Partisan
  asymmetries in online political activity}, EPJ Data Sci.  (2012).
\newblock \doi{10.1140/epjds6}

\bibitem{DelVicario2016c}
M.~{Del Vicario}, G.~Vivaldo, A.~Bessi, F.~Zollo, A.~Scala, G.~Caldarelli,
  W.~Quattrociocchi, {Echo Chambers: Emotional Contagion and Group Polarization
  on Facebook}, Sci. Rep.  (2016).
\newblock \doi{10.1038/srep37825}

\bibitem{DelVicario2017a}
M.~{Del Vicario}, F.~Zollo, G.~Caldarelli, A.~Scala, W.~Quattrociocchi,
  {Mapping social dynamics on Facebook: The Brexit debate}, Soc. Networks
  \textbf{50}, 6 (2017).
\newblock \doi{10.1016/j.socnet.2017.02.002}

\bibitem{Quattrociocchi2014}
W.~Quattrociocchi, G.~Caldarelli, A.~Scala, {Opinion dynamics on interacting
  networks: Media competition and social influence}, Sci. Rep. \textbf{4}
  (2014).
\newblock \doi{10.1038/srep04938}

\bibitem{Zollo2017}
F.~Zollo, A.~Bessi, M.~{Del Vicario}, A.~Scala, G.~Caldarelli, L.~Shekhtman,
  S.~Havlin, W.~Quattrociocchi, {Debunking in a world of tribes}, PLoS One
  \textbf{12}(7) (2017).
\newblock \doi{10.1371/journal.pone.0181821}

\bibitem{Zollo2015}
F.~Zollo, P.K. Novak, M.~{Del Vicario}, A.~Bessi, I.~Mozeti{\v{c}}, A.~Scala,
  G.~Caldarelli, W.~Quattrociocchi, T.~Preis, {Emotional dynamics in the age of
  misinformation}, PLoS One \textbf{10}(9) (2015).
\newblock \doi{10.1371/journal.pone.0138740}

\bibitem{Hentschel2014}
M.~Hentschel, O.~Alonso, S.~Counts, V.~Kandylas, in \emph{International AAAI
  Conference on Web and Social Media} (2014)

\bibitem{varol2019verified}
O.~Varol, I.~Uluturk, Journalists on {T}witter: self-branding, audiences, and
  involvement of bots, Journal of Computational Social Science  (2019)

\bibitem{Saracco2015a}
F.~Saracco, R.~{Di Clemente}, A.~Gabrielli, T.~Squartini, {Randomizing
  bipartite networks: the case of the World Trade Web.}, Scientific Reports
  \textbf{5}, 10595 (2015)

\bibitem{Saracco2016}
F.~Saracco, M.J. Straka, R.~{Di Clemente}, A.~Gabrielli, G.~Caldarelli,
  T.~Squartini, {Inferring monopartite projections of bipartite networks: An
  entropy-based approach}, New J. Phys. \textbf{19}(5), 16 (2017).
\newblock \doi{10.1088/1367-2630/aa6b38}

\bibitem{Blondel2008}
V.D. Blondel, J.L. Guillaume, R.~Lambiotte, E.~Lefebvre, {Fast unfolding of
  communities in large networks}, J. Stat. Mech. Theory Exp.
  \textbf{10008}(10), 6 (2008).
\newblock \doi{10.1088/1742-5468/2008/10/P10008}

\bibitem{Fortunato2010}
S.~Fortunato, {Community detection in graphs}, Phys. Rep. \textbf{486}(3-5), 75
  (2010).
\newblock \doi{10.1016/j.physrep.2009.11.002}

\bibitem{gwet2014}
K.L. Gwet, \emph{Handbook of inter-rater reliability: The definitive guide to
  measuring the extent of agreement among raters} (Advanced Analytics, LLC,
  2014)

\bibitem{DeJeude2018}
J.~van Lidth~de Jeude, R.D. Clemente, G.~Caldarelli, F.~Saracco, T.~Squartini,
  Reconstructing mesoscale network structures, Complexity \textbf{2019},
  5120581:1 (2019)

\bibitem{Raghavan2007b}
U.N. Raghavan, R.~Albert, S.~Kumara, {Near linear time algorithm to detect
  community structures in large-scale networks}, Phys. Rev. E - Stat.
  Nonlinear, Soft Matter Phys.  (2007).
\newblock \doi{10.1103/PhysRevE.76.036106}

\bibitem{Kleinberg1999}
J.M. Kleinberg, {Authoritative sources in a hyperlinked environment}, J. ACM
  (1999).
\newblock \doi{10.1145/324133.324140}

\bibitem{Artime2020}
O.~Artime, V.~D'Andrea, R.~Gallotti, P.L. Sacco, M.~{De Domenico},
  {Effectiveness of dismantling strategies on moderated vs. unmoderated online
  social platforms}, Sci. Rep. \textbf{10}(1), 14392 (2020).
\newblock \doi{10.1038/s41598-020-71231-3}.
\newblock \urlprefix\url{https://doi.org/10.1038/s41598-020-71231-3}

\bibitem{park2004statistical}
J.~Park, M.E.J. Newman, {Statistical mechanics of networks}, Phys. Rev. E
  \textbf{70}(6), 66117 (2004).
\newblock \doi{10.1103/PhysRevE.70.066117}

\bibitem{Garlaschelli2008}
D.~Garlaschelli, M.I. Loffredo, {Maximum likelihood: Extracting unbiased
  information from complex networks}, Phys. Rev. E - Stat. Nonlinear, Soft
  Matter Phys. \textbf{78}(1), 1 (2008).
\newblock \doi{10.1103/PhysRevE.78.015101}

\bibitem{squartini2011analytical}
T.~Squartini, D.~Garlaschelli, {Analytical maximum-likelihood method to detect
  patterns in real networks}, New J. Phys. \textbf{13} (2011).
\newblock \doi{10.1088/1367-2630/13/8/083001}

\bibitem{Diestel2006}
R.~Diestel, \emph{{Graph Theory (Graduate Texts in Mathematics)}} (2006).
\newblock \doi{10.1109/IEMBS.2010.5626521}.
\newblock
  \urlprefix\url{http://www.amazon.com/Graph-Theory-Graduate-Texts-Mathematics/dp/3540261834}

\bibitem{Hong2013}
Y.~Hong, {On computing the distribution function for the Poisson binomial
  distribution}, Comput. Stat. Data Anal. \textbf{59}(1), 41 (2013)

\bibitem{benjamini1995controlling}
Y.~Benjamini, Y.~Hochberg, {Controlling the false discovery rate: a practical
  and powerful approach to multiple testing}, J. R. Stat. Soc. B
  \textbf{57}(1), 289 (1995)

\end{thebibliography}

\end{frontmatter}
\end{document}